\documentclass[final,3p,twocolumn]{elsarticle}

\usepackage{amsmath}
\usepackage{txfonts}
\usepackage[utf8]{inputenc}
\usepackage{graphicx}
\usepackage[protrusion=true,expansion=true]{microtype}
\usepackage{hyperref,xcolor}%lineno,
\usepackage[T1]{fontenc} % if needed
\usepackage{graphicx}
\usepackage{ulem}

\usepackage{amssymb} % collection de symboles mathématiques
\usepackage{tabularx} % gestion avancée des tableaux
\usepackage{gensymb}
\usepackage{appendix}
\usepackage{enumerate}   
\usepackage{bm}
\usepackage{setspace}
\usepackage{textcomp}

\journal{Astroparticle Physics}

\begin{document}

\begin{frontmatter}

\title{Polarisation signatures in radio\\ for inclined cosmic-ray induced air-shower identification}

\author[a]{Simon Chiche}
\ead{simon.chiche@iap.fr}
\author[a,f]{Kumiko Kotera}
\author[b,a,e]{Olivier Martineau-Huynh}
\author[c,d,a]{Matias Tueros}
\author[f]{Krijn D. de Vries}

\address[a]{Sorbonne Universit\'{e}, CNRS, UMR 7095, Institut d'Astrophysique de Paris, 98 bis bd Arago, 75014 Paris, France}
\address[b]{Sorbonne Université, Université Paris Diderot, Sorbonne Paris Cité, CNRS,Laboratoire de Physique Nucléaire et de Hautes Energies (LPNHE), Paris, France}
\address[c]{IFLP - CCT La Plata - CONICET, Casilla de Correo 727 (1900) La Plata, Argentina}
\address[d]{Depto. de Fisica, Fac. de Cs. Ex., Universidad Nacional de La Plata,  Casilla de Coreo 67 (1900) La Plata, Argentina}
\address[e]{National Astronomical Observatories, Chinese Academy of Sciences, Beijing 100012, China}
\address[f]{Vrije Universiteit Brussel, Physics Department, Pleinlaan 2, 1050 Brussels, Belgium}

\date{Received $\langle$date$\rangle$ / Accepted $\langle$date$\rangle$}

\begin{abstract}
Autonomous radio-detection, i.e., detection of air-showers with standalone radio arrays, is one of the major technical challenges to overcome for the next generation astroparticle detectors.
In this context, we study polarisation signatures of simulated radio signals to perform an identification of the associated air-showers initiated by cosmic-rays and neutrinos. We compare the two sources of radio emission (the charge excess and geomagnetic) and show that the former is almost negligible for inclined (zenith angle $>65\degree$) cosmic-ray air-showers. This provides an efficient background rejection criterion at the DAQ level, based on the projection of the total electric field along the direction of the local magnetic field. This relevant quantity  can be computed, ---even in an online treatment--- for antennas measuring three orthogonal polarisations.
Independently of the experimental antenna layout, we estimate that assuming a random polarisation of noise events, a rejection from $\approx 72\%$ (for a non favorable detector location) to $\approx 93\%$ (for a favorable one) of the noise induced events and a trigger efficiency of 86\% (93\%) with a $3\sigma$ ($5\sigma$) trigger threshold level should be achievable. We also show that neutrino-induced showers present a charge excess to geomagnetic signal ratio up to $\sim 10$ times higher than for cosmic ray showers. Although this characteristic makes the identification of neutrino-induced showers challenging via the method developed here, it provides an efficient criterion to perform an offline discrimination between cosmic-ray and neutrino primaries. The stronger charge excess emission will also help the reconstruction of  air-shower parameters, such as the core position.

\end{abstract}

\begin{keyword}
radio-detection, high-energy astroparticles, air-shower simulation, polarisation
\vspace{0.5cm}
\noindent

\end{keyword}

\end{frontmatter}

%\linenumbers

\newpage

\newcommand{\red}{\color{red}}

%%%%%%%%%%%%%%%%%%%%%%%%%%%%%%%%%%%%%%%%%%
%%%% INTRODUCTION
%%%%%%%%%%%%%%%%%%%%%%%%%%%%%%%%%%%%%%%%%%

\section{Introduction}

Thanks to important experimental and theoretical progress in the past decade, radio-detection was established as a robust and efficient technique to detect and reconstruct the parameters of cosmic-ray induced air-showers (e.g., \cite{Huege:2016veh,2017PrPNP..93....1S} for reviews). So far, most of these successful radio experiments have combined radio antennas with particle detector arrays (e.g., AERA \cite{Abreu2012AntennasFT}, CODALEMA \cite{Carduner:2017mqf}, LOFAR \cite{Corstanje:2014waa}, Tunka-Rex \cite{Prosin:2016jev}).

{\it Autonomous} radio-detection however, i.e., identifying radio signals with radio antennas alone (without an external trigger such as particle detectors) is a real challenge. The difficulties stem from the fact that transient pulses induced by background sources (radio relays, planes, storms...) are far more numerous -- by a factor $10^{4}$ at least outside polar areas \cite{2011APh....34..717A} -- than those induced by astroparticles. This requires an efficient rejection process performed as early as possible in the experimental chain. A handful of experiments have already shown promising results towards autonomous radio detection~\cite{Charrier:2018fle,Hoover2010ObservationOU,Barwick:2016mxm,Romero-Wolf:20193f,Huege_2019}, yet more robust methods of background rejection have to be developed to achieve higher efficiency. Autonomous radio-detection is of primary importance for the next generation of radio experiments with a large number of antennas, such as GRAND \cite{2020SCPMA..6319501A}. 

Most of the experiments cited above have mostly detected cosmic-ray air-showers with zenith angles lower than 60{\degree}.  Very inclined air-showers were also studied by ANITA and AERA for example and present different properties due to their longer propagation in a thiner atmosphere. The detection and study of such events has gained momentum, with the conception of large-scale and sparse radio arrays (e.g., GRAND \cite{2020SCPMA..6319501A} and the Auger Radio Upgrade \cite{Aab:2016vlz}), endeavouring to sample the large radio footprints of inclined showers. Such inclined showers could be generated by ultra-high-energy tau neutrinos with Earth-skimming trajectories.

In this study, we will investigate the specific polarisation features of the radio signal and establish an identification procedure which provides an instrumental way to address the challenge of autonomous radio-detection for cosmic-ray induced inclined air-showers. Additionally, we will also show how polarisation signatures could enable the discrimination of cosmic-ray and neutrino induced air-showers.
 
 The polarisation is defined as the direction of the electric field vector at a given instant. As the radiation from particles in the shower adds up coherently, we expect a highly polarised signal~\cite{Schellart:2014oaa,Scholten:2017tcr}. The emission mechanisms that contribute to the polarisation are well known and were overly described by Classical Electrodynamics calculations~\cite{2012NIMPA.662S..80S,2008APh....29...94S,2010APh....34..267D}. Polarisation can directly be measured with radio antennas with two or three arms, that would measure the different components of the electric field. Polarisation is known to be an efficient tool to perform background rejection \cite{Schellart:2014oaa, 2017arXiv170601451G, huege2019symmetrizing,Aab:2014esa, Aab_2016_2}, we aim here at quantifying quantitatively how efficient this tool can be for the detection of very inclined air-showers.

We recall the main mechanisms responsible for the radio emission in air-showers in Section~\ref{section:radio_emission}. In Sections~\ref{section:zhaires} and \ref{section:polarisation_reconstruction}, we reconstruct the polarisation from numerical simulations of air-showers. Then in Section~\ref{section:ce_geo_ratio}, we use the reconstructed polarisation to highlight the characteristics that we exploit to perform autonomous radio detection. We present our results in Section~\ref{section:identification_method}, first on a star-shape detector layout and then on an experimental layout and define criteria for trigger at the DAQ level. Finally, in Section~\ref{section:neutrino}, we examine the case of neutrino induced air-showers.

%%%%%%%%%%%%%%%%%%%%%%%%%%%%%%%%%%%%%%%%%%
%%%% THE METHOD - an overview
%%%%%%%%%%%%%%%%%%%%%%%%%%%%%%%%%%%%%%%%%%
\section{Radio emission from extensive
air-showers}\label{section:radio_emission}

The radio emission from extensive air-showers is mainly the superposition of two mechanisms with different linear polarisation patterns: the geomagnetic emission and the charge excess emission. 

\subsection{Geomagnetic emission}\label{section:geomagnetic}

\begin{figure*}[tb]
\includegraphics[width=0.49\linewidth]{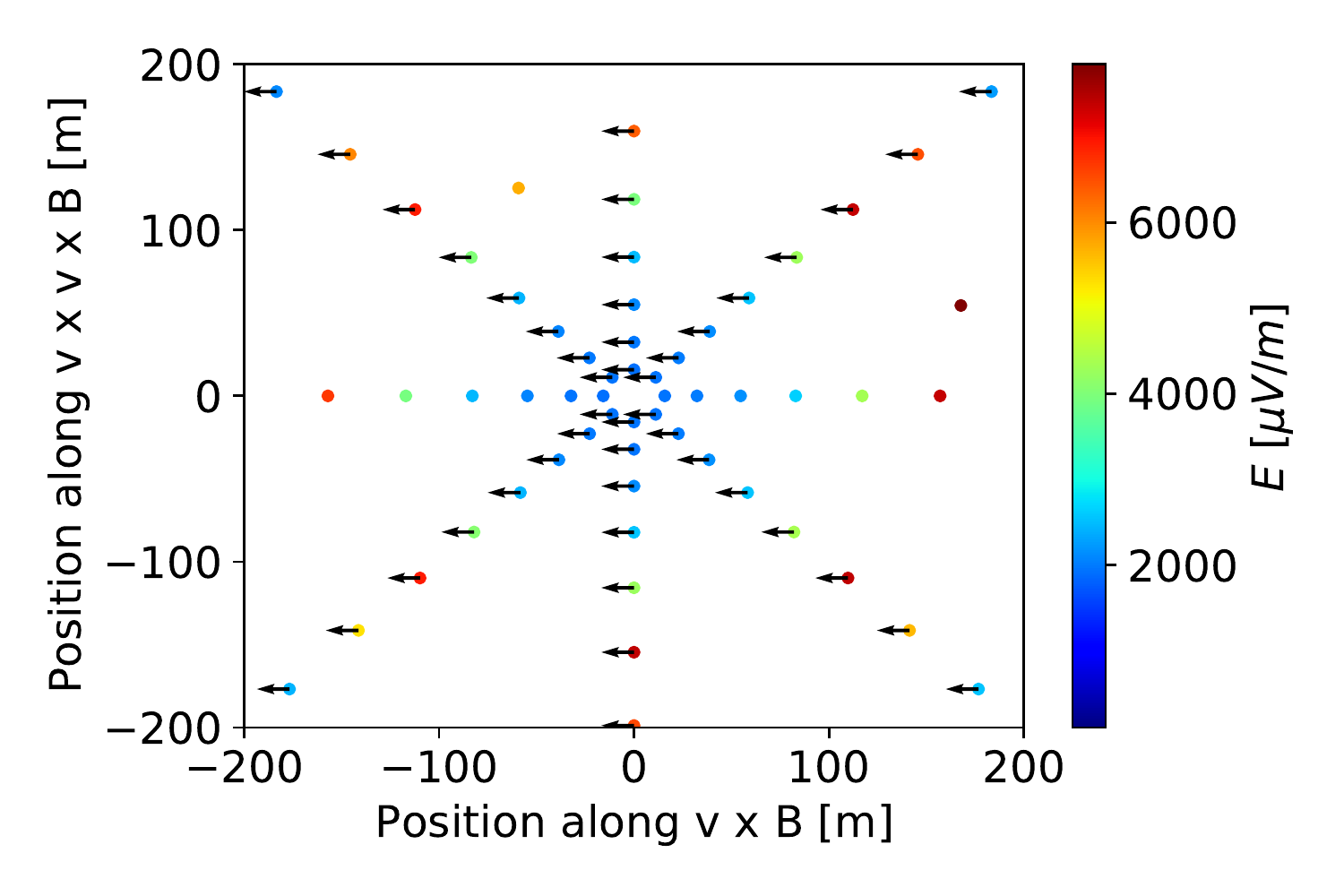}\hfill
\includegraphics[width=0.49\linewidth]{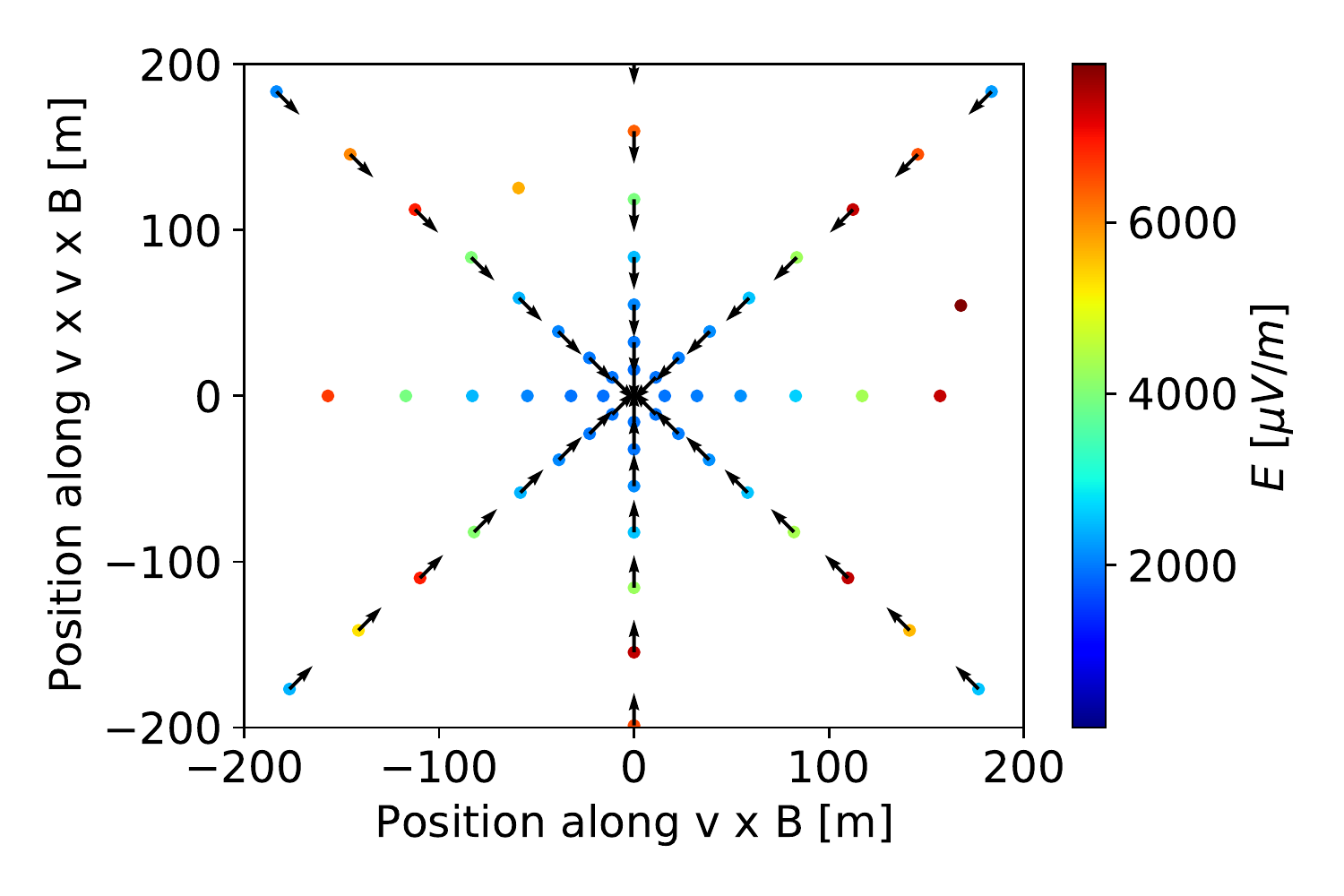}
\caption{Total (geomagnetic + charge excess) electric field amplitude (color scale) and direction (arrows) of the geomagnetic ({\it left}) and charge excess ({\it right}) emissions, for a shower with primary particle energy $E=0.68\,$EeV and zenith angle $\theta=53$\degree, represented in the plane perpendicular to the shower axis (${\bf e}_{\mathbf{v}\times \mathbf{B}}$, ${\bf e}_{\mathbf{v}\times (\mathbf{v}\times \mathbf{B})}$). A small number of antennas is also generated outside of the star-shape pattern in our simulations, for cross checks.}\label{fig:polarisation}
\end{figure*}

In an Extensive Air-Shower (EAS), particles of charge $q$ are deflected by the geomagnetic field $\mathbf{B}$ due to the Lorentz force:
$\mathbf{F} = q\mathbf{v}\times \mathbf{B}$ where $\mathbf{v}$ is the particle velocity. The deflection in opposite direction of the lightest particles mostly, i.e., electrons and positrons in combination with an effective friction force due to multiple scattering results in a net transverse current in the cascade front. This current varies over time as the number of electrons and positrons changes during shower development. This induces a linearly polarised radio emission orthogonal to the direction of the Earth  magnetic field and the shower axis, i.e., along $-\mathbf{v}\times \mathbf{B}$ as can be seen on the left panel of Fig.~\ref{fig:polarisation}. 

The geomagnetic electric field amplitude scales with $\approx vB\sin{\alpha}$, with $\alpha$ the angle between the shower axis and the local Earth magnetic field direction. It also increases with the number of charged particles (coherence effect) and with the mean free path of electrons before they loose their energy during strong inelastic processes like Bremsstrahlung. This implies that the amplitude of the geomagnetic emission is expected to decrease with increasing density of the medium.

\subsection{Charge excess or Askaryan emission}\label{section:charge_excess}

As the shower propagates through the atmosphere, electrons from the air atoms are struck by high energy shower particles. This results in an accumulation of negative charges in the shower front, mostly close to the intersection of the shower axis and the shower front, the so-called shower core. The net negative charges can be assimilated to a point charge at the shower core comparatively to the longitudinal extent of the shower. This results in an electric field oriented toward the core. As illustrated in the right panel of Figure~\ref{fig:polarisation}, the charge excess emission leads to a linear radial polarisation in the shower plane $({\bf e}_{\mathbf{v}\times \mathbf{B}}, {\bf e}_{\mathbf{v}\times (\mathbf{v}\times \mathbf{B})})$, defined as the plane perpendicular to the shower axis, with ${\bf e}_{\mathbf{v}\times \mathbf{B}}$ and ${\bf e}_{\mathbf{v}\times (\mathbf{v}\times \mathbf{B})}$, unit vectors along the $\mathbf{v}\times \mathbf{B}$ and $\mathbf{v}\times (\mathbf{v} \times \mathbf{B})$ directions respectively. 

The charge excess electric field amplitude increases with the density of the medium in which the shower develops~\cite{2017PrPNP..93....1S}. It accounts for about $10\%$ of the geomagnetic amplitude for vertical air-showers after correction of $\sin{\alpha}$.

\section{ZHAireS simulations and signal processing}\label{section:zhaires}

This study is based on the analysis of data obtained with the ZHAireS code. ZHAireS is a Monte-Carlo simulation of the microscopic emission by individual particles in the shower~\cite{2012APh....35..325A, ZHAireS}. These simulations can generate showers with various primary particles (proton, iron, gamma...), various directions (azimuth and zenith angles) and various energies. We used the 19.04.00 version of Aires with SYBILL2.3 as hadronic model combined with the 1.0.28 version of the ZHAireS extension. All the simulations performed for this article were done using a magnetic field inclination of $60.79\degree$, while the observer altitude level for star-shape simulations was fixed at 1080\,m above sea level.

\begin{figure*}[tb]
\includegraphics[width=0.49\linewidth]{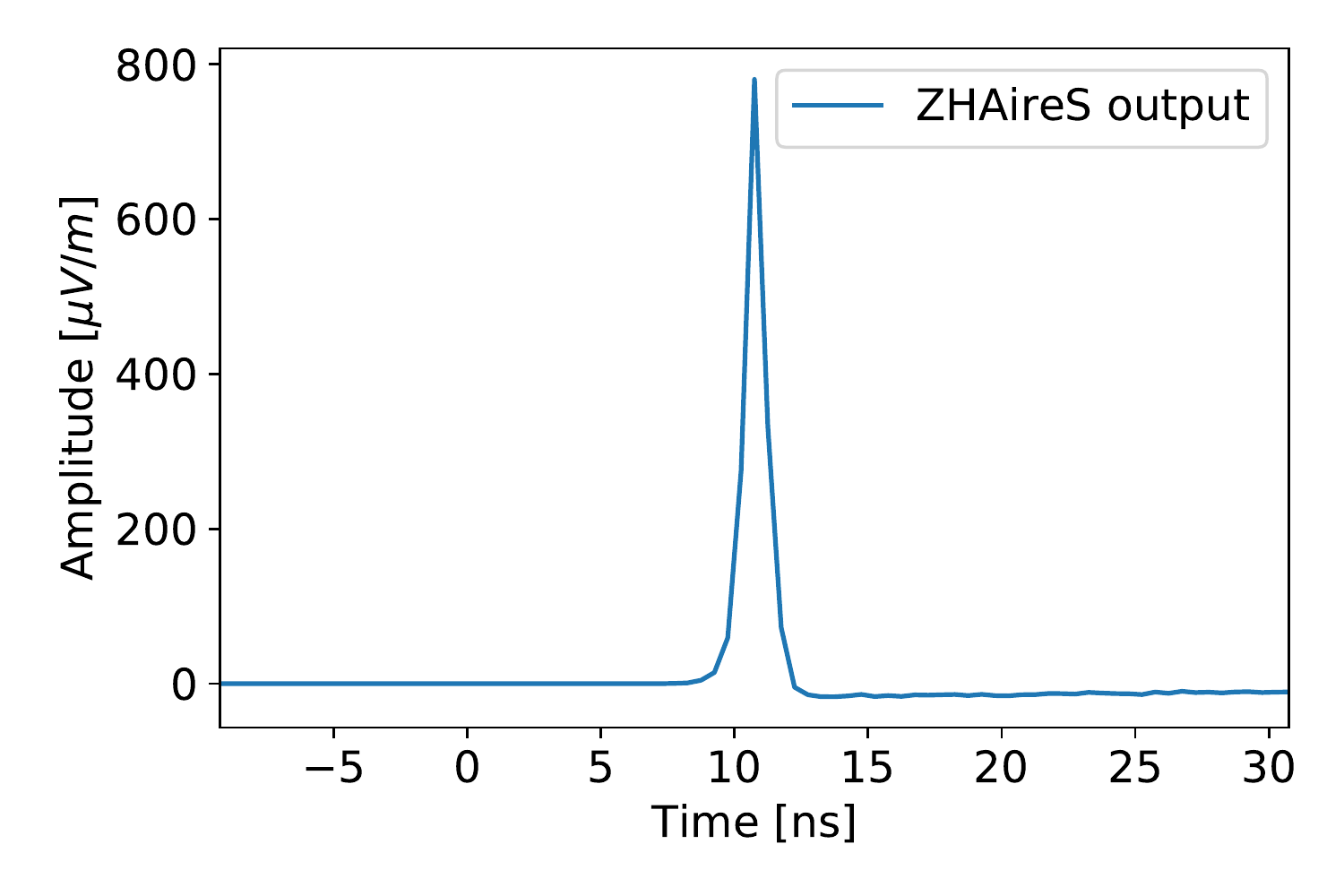}\hfill
\includegraphics[width=0.49\linewidth]{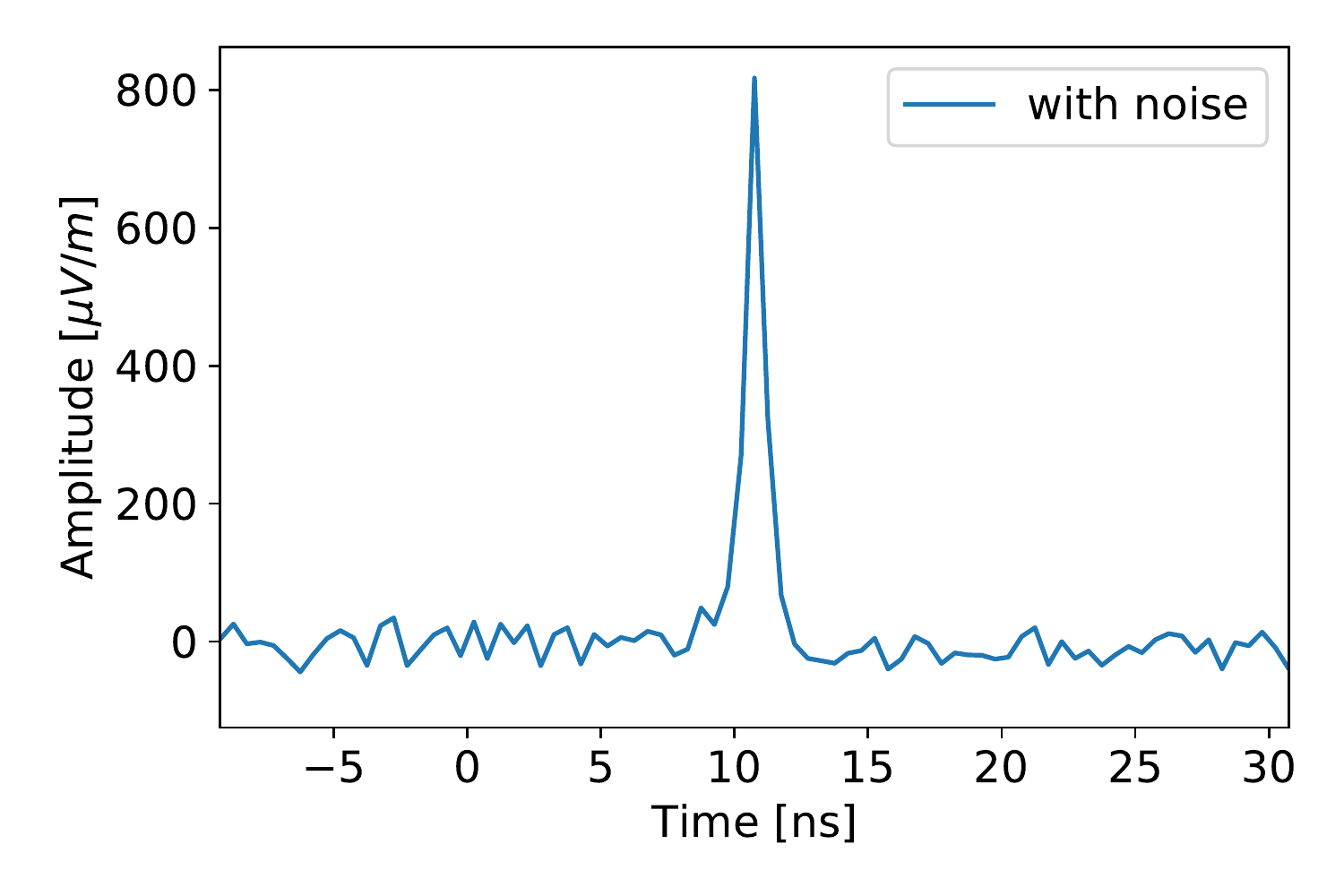}
\caption{({\it Left})  Example of a raw trace along the $y$-axis (East-West) at a given antenna, for a shower with primary energy $E = 0.68\,$EeV and zenith angle $\theta = 53\degree$. ({\it Right}) same trace after adding a gaussian stationary noise of $RMS = 20\mu$V/m. }\label{fig:traces_geo}
\end{figure*}

\begin{figure}[tb]
\centering 
\includegraphics[width=0.95\columnwidth]{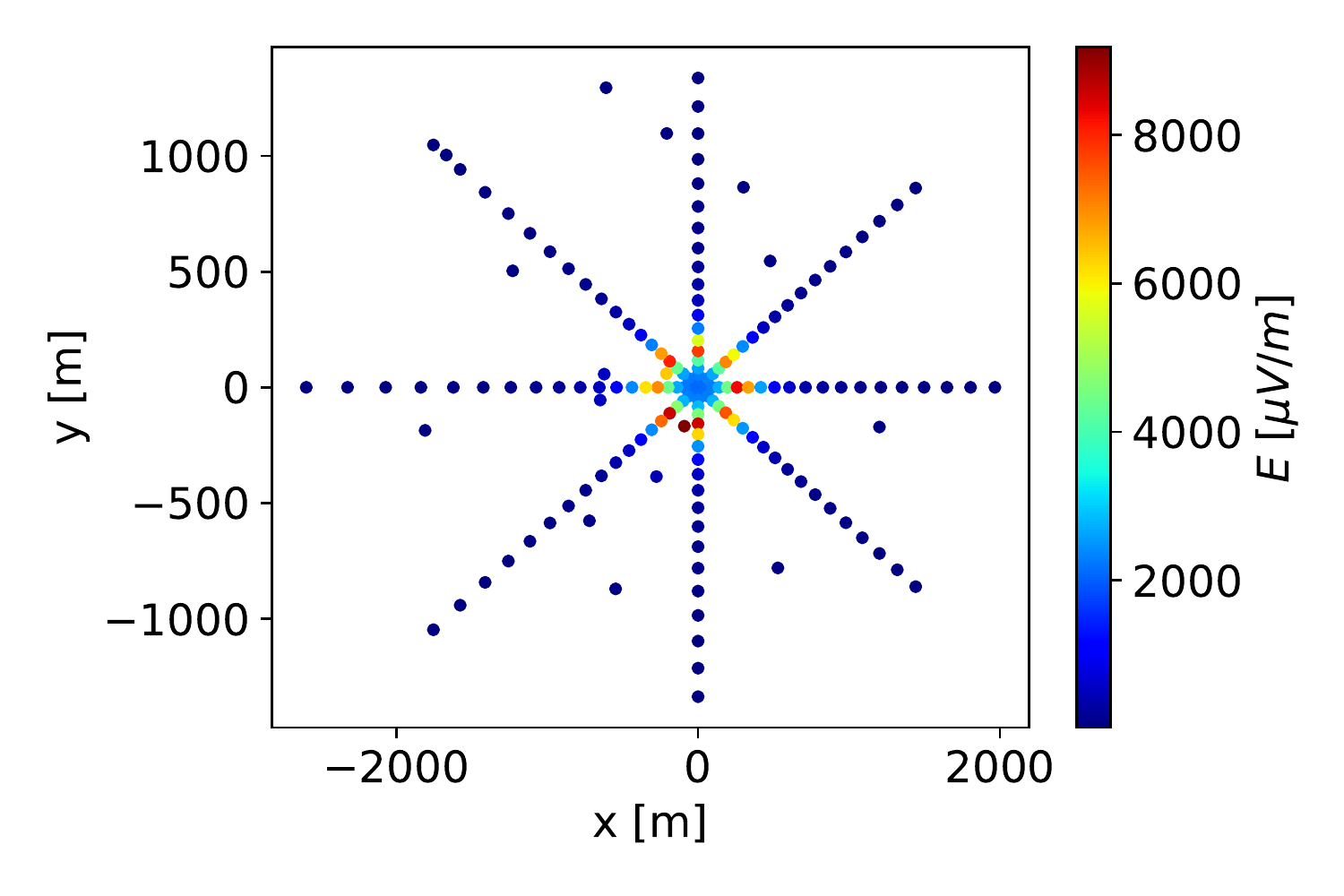}
\vspace{-0.4cm}
\caption{Total electric field peak-to-peak in the geographic plane for a star-shape antenna layout considering a proton induced shower with primary particle energy $E = 0.68\,$EeV and zenith angle $\theta = 53\degree$.}\label{fig:p2p_sp}
\vspace{-0.4cm}
\end{figure}

The outputs of ZHAireS simulations are the electric field time-traces $E_{\rm x}(t)$, $E_{\rm y}(t)$, $E_{\rm z}(t)$ measured at each antenna position, where we choose $x$ = South-North, $y$ = West-East and $z$ = up. Figure~\ref{fig:traces_geo} (left panel) shows an example of $E_{\rm y}(t)$ received at one given antenna and Figure~\ref{fig:p2p_sp} the total peak-to-peak amplitude for a star-shape layout of antennas. The trace is sampled at a 0.5 ns rate. The layout here is centered on the air-shower core position, it appears that the signal is peaked on a circular region which corresponds to antennas on the Cerenkov cone of the radio signal. 

To reconstruct the polarisation, we model the experimental measurement of the radio signal by the antennas. For this purpose, we first add a background noise to the traces to model the stationary noise coming from the Milky Way.  This is done by adding a Gaussian stationary noise of root mean square (RMS) 20\,$\mu$V/m,  computed following Eq.~2 of~\cite{Decoene_2021}, where the sky brightness follows the frequency dependency computed from Galactic emission measurements. This is illustrated in the right panel of Fig.~\ref{fig:traces_geo}. We consider that there is no correlation between the three components ($x$, $y$ and $z$) of the noise and therefore add a different noise realisation to each channel of the traces. It should be noted that ideally this study should be repeated on actual noise measurements for example at the GRAND site. However, with no such data available, we performed an idealized study using randomly polarized Gaussian noise.

\begin{figure}[tb]
\centering 
\vspace{-0.4cm}
\includegraphics[width=0.95\columnwidth]{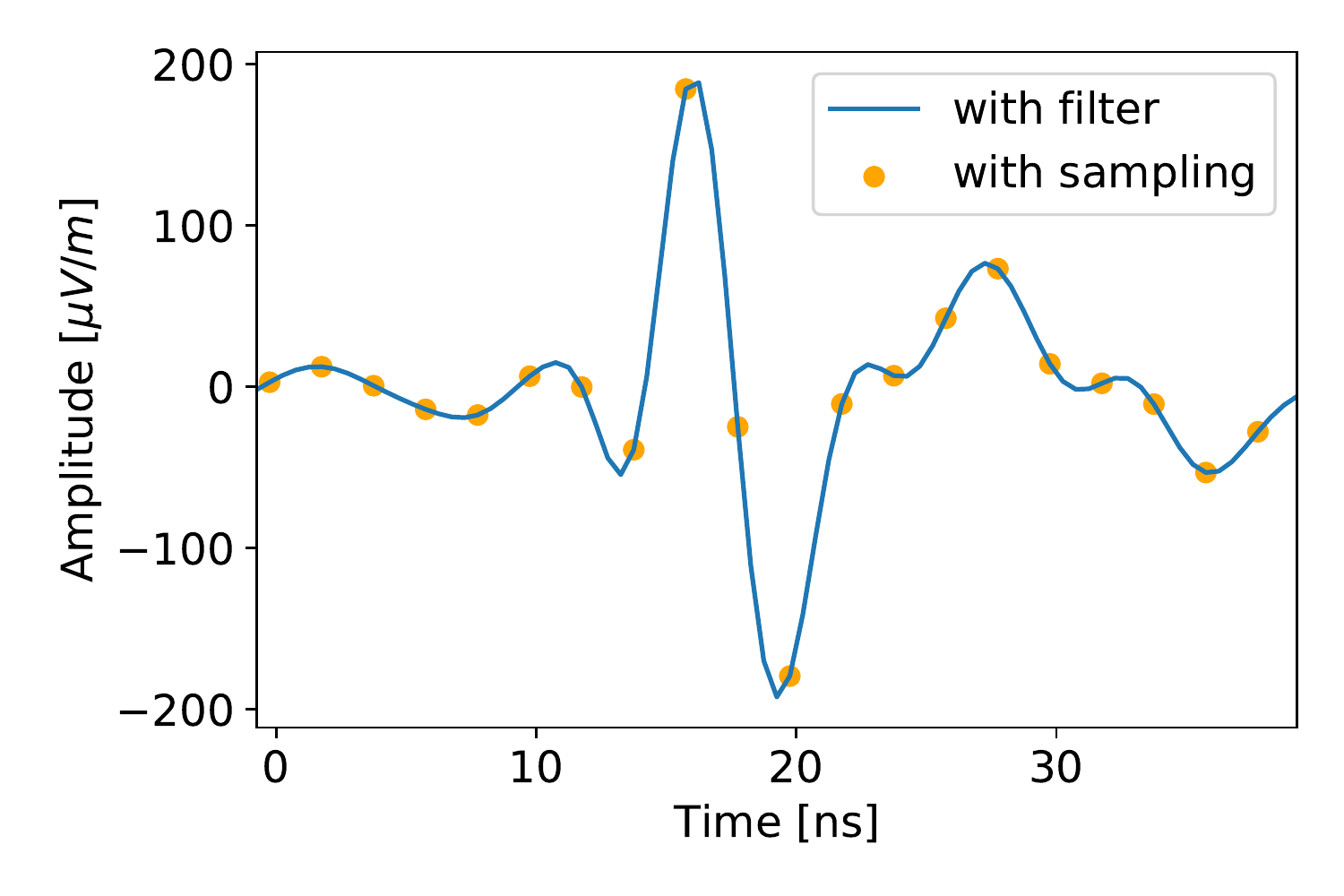}
\caption{Traces from Fig.~\ref{fig:traces_geo} after filtering (solid line), and sampling (orange dots).}\label{fig:filter_sample}
\end{figure}

We then apply a Butterworth filter of order 5 in the frequency range of the GRAND experiment ($50-200$\,MHz). Finally, we sample the traces at a 2\,ns period to derive the experimental measurement of the radio signal. The filtering and the sampling processes are illustrated in Fig.~\ref{fig:filter_sample}.

Once the traces are fully processed (noise, filtering, sampling), we can reconstruct the polarisation.
As we aim to present general features of the polarisation independent of the experiment, we do not take into account any specific antenna response in this treatment. However, to account for the calibration error when reconstructing the polarisation, we apply a factor randomly drawn in a Gaussian distribution of mean value 1.0 and standard deviation 0.1 for each antenna. An identical value is chosen for the $x$ and $y$ axes, and an independent one for the $z$ axis, thus accounting for the different antenna response and wave reflection for horizontal and vertical polarisations.

\section{Polarisation reconstruction}
\label{section:polarisation_reconstruction}

We detail here the steps to reconstruct the polarisation from the time traces presented in the previous section. The analysis presented in this Section and in Section~\ref{section:ce_geo_ratio} follows mainly what was done by the Auger collaboration in~\cite{Aab_2016}, but is presented here with our own set of simulations for self-consistency. We use a standard method based on the Stokes parameters as detailed in Ref.~\cite{Aab:2014esa, Schoorlemmer:2012xpa}, and work in the shower plane (${\bf e}_{\mathbf{v}\times \mathbf{B}}$, ${\bf e}_{\mathbf{v}\times (\mathbf{v}\times \mathbf{B})}$), the relevant framework to study polarisation.

\subsection{Stokes parameters}\label{section:Stokes_parameters}
The Stokes parameters are four parameters that allow to completely describe the polarisation of a temporal signal in a plane. For temporal values of the electric field $E_{v \times B}(t)$, $E_{v \times (v \times B)}(t)$ given in a plane, the Stokes parameters are defined as follows:
\begin{eqnarray}
I &=& \dfrac{1}{n}\sum_{i = 1}^{n}x_{i}^{2} + \widehat{x}_{i}^{2} + y_{i}^{2} + \widehat{y}_{i}^{2} 
\label{equation: I_stokes}\\
Q &=& \dfrac{1}{n}\sum_{i = 1}^{n}x_{i}^{2} + \widehat{x}_{i}^{2} - y_{i}^{2} - \widehat{y}_{i}^{2} \\
U &=& \dfrac{2}{n}\sum_{i = 1}^{n} x_{i}{y}_{i} + \widehat{x}_{i}\widehat{y}_{i}\\
V &=& \dfrac{2}{n}\sum_{i = 1}^{n} \widehat{x}_{i}{y}_{i} - {x}_{i}\widehat{y}_{i}\ ,
\end{eqnarray}
where $n$ corresponds to the number of time sample used for computation, $x_{i}$ and $y_{i}$ correspond to the values of the traces $E_{v \times B}(t)$, $E_{v \times (v \times B)}(t)$ at an instant $i$ and $ \widehat{x}_{i}$ and $\widehat{y}_{i}$ to the imaginary part of the traces obtained by extending the traces in the complex domain using the Hilbert transform.

\begin{figure}[tb]
\centering 
\includegraphics[width=0.95\columnwidth]{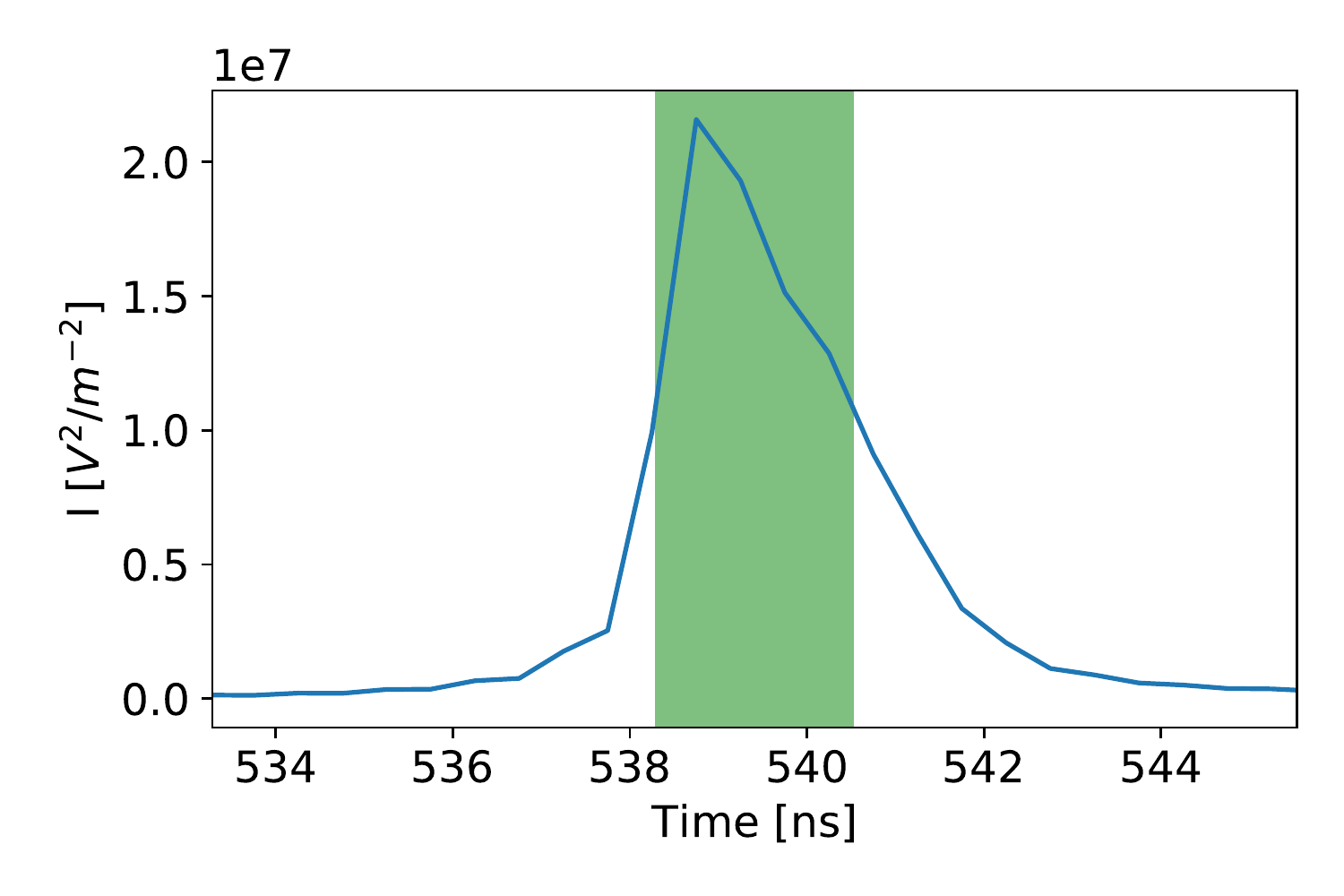}
\caption{Example of time window defined using the FWHM of the peak for a shower with $E = 0.68\,$EeV and $\theta = 53\degree$.}\label{fig:time_window}
\end{figure}

The Stokes parameter $I$ is related to the total intensity of the radio signal and can hence be used to define the time window for the computation of Stokes parameters. Instead of averaging $I$ over the full time sample as in Eq.~\ref{equation: I_stokes}, we can construct a time series and derive the intensity for each individual sample. We then define the time window over which the traces and the Stokes parameters are averaged as the FWHM\footnote{The full width half maximum corresponds to the with of the peak at half of its maximum.} of the time series as was done by Ref.~\cite{Schoorlemmer:2012xpa} and illustrated in Figure~\ref{fig:time_window} for a given trace. This time window is designed to maximise the signal-to-noise ratio of the signal.

\subsection{Polarisation in the shower plane}\label{section:polarisation shower plane}

For any elliptical polarisation, the Stokes parameters $U$ and $Q$ can be expressed as a function of the ellipticity angles $\Psi$ and $\chi$ as $Q =I\cos{2\psi}\cos{2\chi}$ and $U =I\cos{2\psi}\sin{2\chi}$. In the particular case of a linear polarisation and for Stokes parameters computed in the shower plane we  have $\chi = \phi_{\rm p}$, where $\phi_{\rm p}$ is the polarisation angle, i.e., the angle between the direction of the polarisation and the ${\bf v} \times {\bf B}$ direction, as illustrated in Fig.~\ref{fig:phi_p_sketch}. Then, once we have averaged the 4 Stokes parameters over the time window displayed in Fig.~\ref{fig:time_window} the polarisation angle reads
\begin{equation}
\phi_{\rm p} = 0.5\arctan{\frac{U}{Q}} \ .
\end{equation}

Assuming the total polarisation to be linear, we can finally express the components of the polarisation in the shower plane (directly related to the direction of the electric field vector) as follows:
\begin{eqnarray}
    E_{v \times B} &=& \sqrt{I}\cos{\phi_{\rm p}} \ , \label{eq:Evxb}\\
    E_{v \times (v\times B)} &=& \sqrt{I}\sin{\phi_{\rm p}} \label{eq:Evxvxb} \ .
\end{eqnarray}

\begin{figure}[tb]
\centering
\includegraphics[width=0.95\columnwidth]{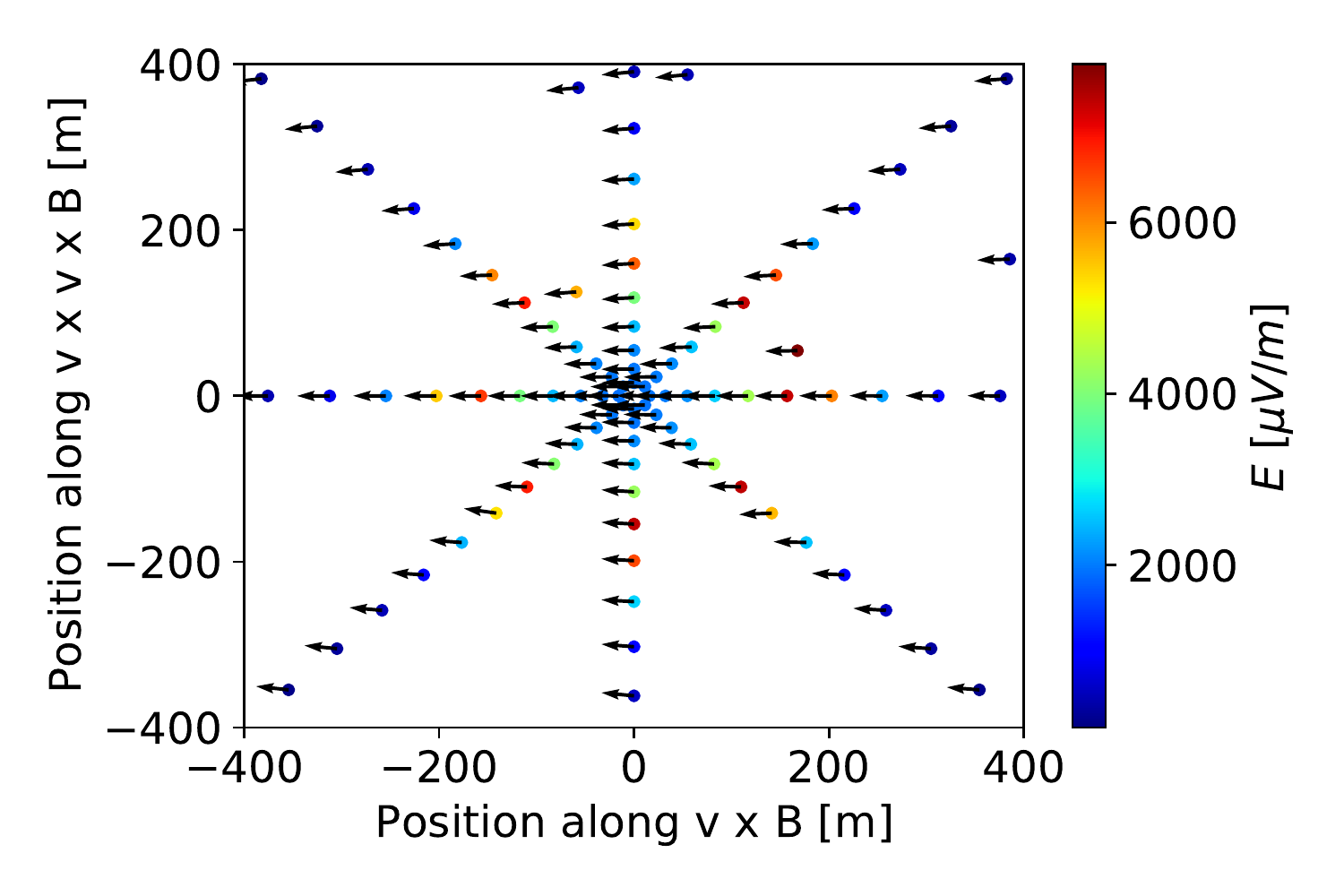}
\caption{Total polarisation inferred from Stokes parameters in the shower plane for a shower with primary particle energy $E = 0.68\,$EeV and zenith angle $\theta = 53\degree$.}\label{fig:polarisation_stokes}
\end{figure}

\begin{figure}[tb]
\centering 
\includegraphics[width=0.95\columnwidth]{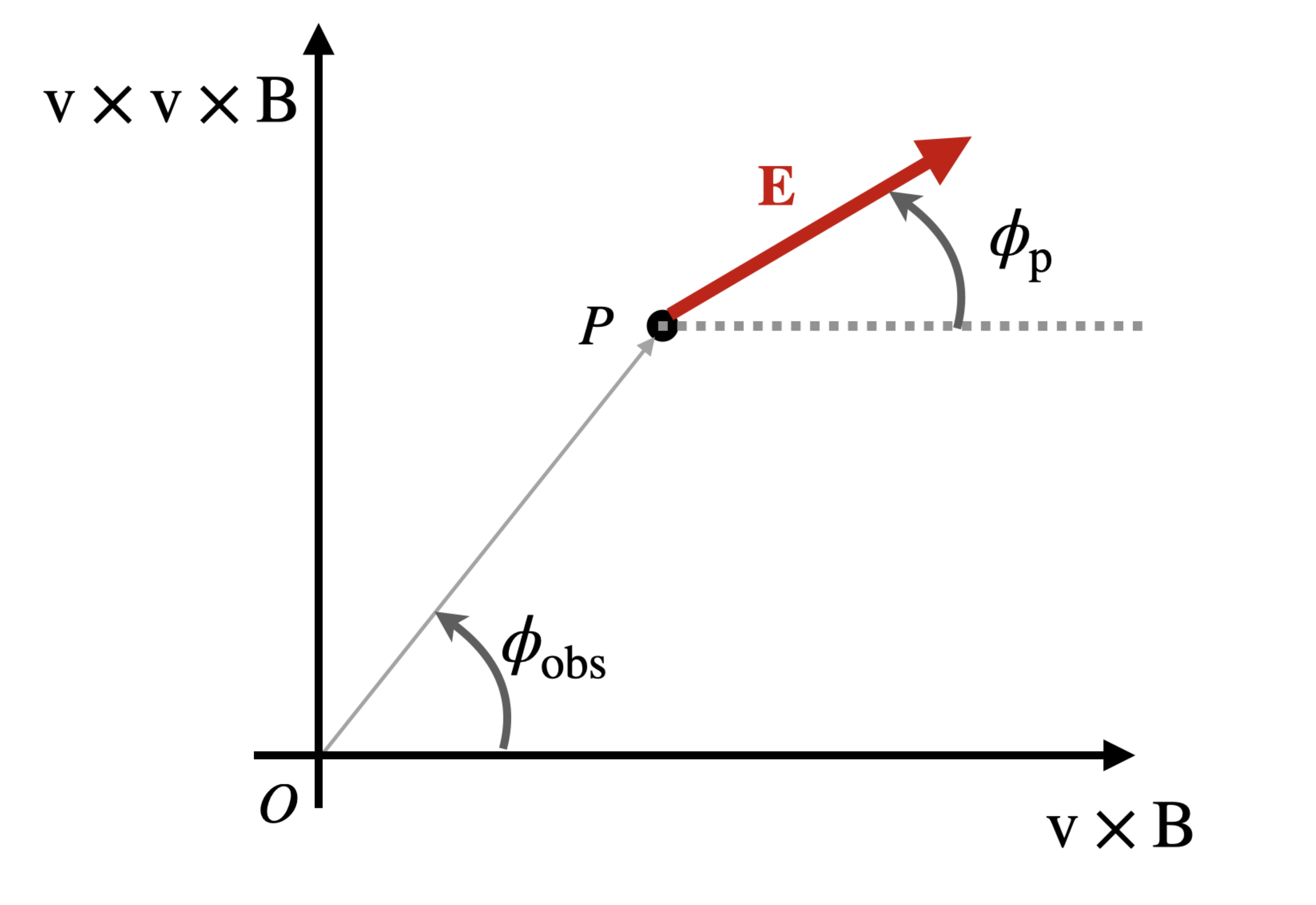}
\caption{Representation of $\phi_{\rm p}$ and $\phi_{\rm obs}$ in the shower-plane frame $(O,{\bf e}_{{\bf v}\times {\bf B}},{\bf e}_{{\bf v}\times{\bf v}\times {\bf B}})$, with $O$ the shower core position, ${\bf B}$ the geomagnetic field and ${\bf k}$ the shower axis. ${\bf E}$ is the electric field at observer position $P$.} \label{fig:phi_p_sketch}
\end{figure}

The reconstructed polarisation in the shower plane is represented in Fig.~\ref{fig:polarisation_stokes} for a given shower. One can observe that the total polarisation is predominantly aligned with $- \mathbf{v \times B}$ as expected for a dominant geomagnetic emission. 

\section{Characteristics of the charge excess to geomagnetic ratio}\label{section:ce_geo_ratio}

Exploiting polarisation patterns of both emission mechanisms (a geomagnetic contribution along $-\mathbf{v}\times \mathbf{B}$ and a radial charge excess), we can use the total polarisation to infer the amplitude of each mechanism. In particular, as the field along $\mathbf{v}\times (\mathbf{v} \times \mathbf{B})$ corresponds to charge excess only, knowing the direction of the charge excess emission (radial in the shower plane), we can infer its norm. The field along $\mathbf{v} \times \mathbf{B}$ however, is the result of the interference between the charge excess and geomagnetic emission. Therefore after inferring the norm of the charge excess we can subtract its contribution to the $\mathbf{v} \times \mathbf{B}$  component of the electric field to derive the norm of the geomagnetic emission. Hence, amplitudes of both mechanisms can be derived from the following expressions \cite{huege2019symmetrizing,Aab_2016}
\begin{eqnarray}
    E_{\rm ce} &=& \frac{E_{\rm v \times (v \times B) }}{|\sin{\phi_{\rm obs}}|} \ , \label{eq:norm_ce}\\ 
    E_{\rm  geo} &=& E_{\rm v \times B}  - E_{\rm v \times (v \times B)}\frac{\cos{\phi_{\rm obs}}}{|\sin{\phi_{\rm obs}}|} \ , \label{eq:norm_geo}
\end{eqnarray}
where $\phi_{\rm obs}$ is the angle between the antenna position and the $\mathbf{v}\times \mathbf{B}$ axis. One should note that these equations are not applicable for the horizontal baseline of antennas (along the $\mathbf{v}\times \mathbf{B}$ axis) as along this baseline, the charge excess has no component along $\mathbf{v} \times (\mathbf{v} \times \mathbf{B})$. \\

From the reconstructed components of the charge excess and geomagnetic emission, we can then infer the charge excess to geomagnetic ratio defined as follows:
\begin{equation}
a\equiv \sin \alpha \, \frac{E_{\rm ce}}{E_{\rm geo}} \ , \label{eq:ce_geo}
\end{equation} 
where we correct the geomagnetic emission from $\sin{\alpha}$. This is done in order to infer general results that do not depend on the azimuth of the shower. \\

In the following sections, we will present the dependencies of the $a$ ratio with shower parameters. Following what was presented in~\cite{Aab_2016_2}, we will study how this ratio provides an efficient way to identify signatures from inclined cosmic-ray air-showers and allows to impose constraints on the trigger condition to perform a strong background rejection. We will also demonstrate that it can help discriminate between cosmic-ray and neutrino induced showers.

\subsection{Evolution with angular distance}

\begin{figure}[tb]
\centering 
\includegraphics[width=0.95\columnwidth]{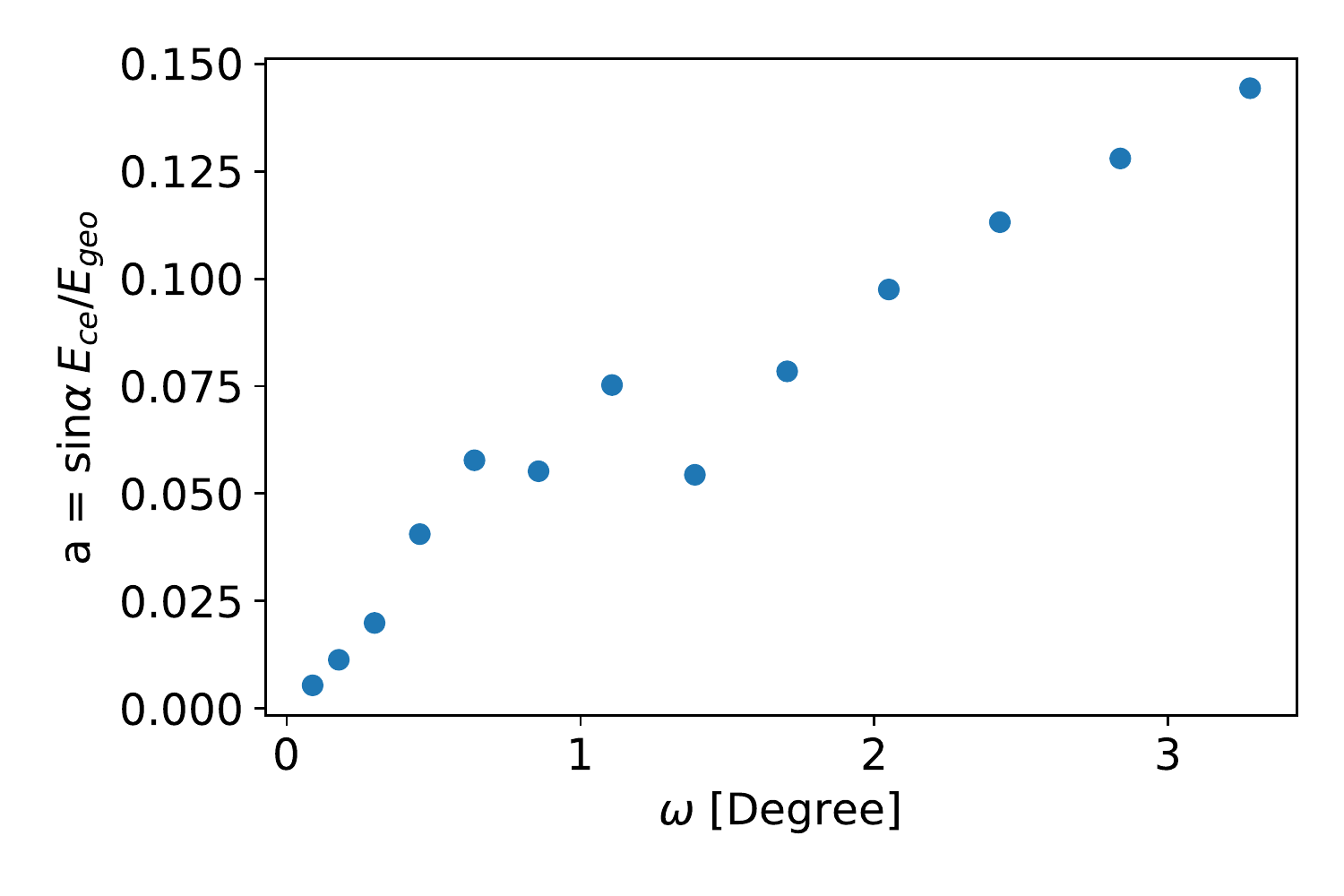}
\caption{Evolution with angular distance $\omega$ of the charge excess to geomagnetic ratio as reconstructed with Eqs.~(\ref{eq:norm_geo}-\ref{eq:norm_ce}) for a simulation of a proton-induced shower with energy $E = 0.68\,$EeV and zenith angle $\theta = 53\degree$ on a star-shape layout.}\label{fig:ce_geo_w_starshape}
\end{figure}

In Figure~\ref{fig:ce_geo_w_starshape}, we represent the charge excess to geomagnetic ratio for a given simulation without noise on a star-shape layout as a function of $\omega$, the angular deviation to the shower axis measured from $X_{\rm max}$, i.e. ($\widehat{\bf{u_{v}}, \bf{u_{\rm antenna}}}$), with $\bf{u_{v}}$ the unit vector related to the direction of the shower and $\bf{u_{\rm antenna}}$ the unit vector that goes from $X_{\rm max}$ to a given antenna. Using this angle is particularly convenient as shower properties are expected to be similar for antennas located at the same $\omega$. Indeed, the amplitude pattern of the radio signal in the shower plane becomes invariant by translation along the shower axis with this representation.  We can observe a quasi-linear increase of the charge excess to geomagnetic ratio with $\omega$. This linear increase with $\omega$ ---or similarly with distance to the shower core--- was already discussed in~\cite{2013APh....45...23D} or~\cite{Schellart:2014oaa} and can be justified considering simple scaling laws within the analytical framework of Ref.~\cite{Scholten:2017tcr}. As the charge excess contributes to the scalar potential and the geomagnetic to the vector potential~\cite{2008APh....29...94S}, the related electric fields read
\begin{eqnarray}
    E_{\rm geo} &\approx& \frac{{\rm d}t_{r}}{{\rm d}t}\frac{{\rm d}A^{\rm geo}}{{\rm d}t_{r}} \ ,  \\
    E_{\rm ce} &\approx& \frac{{\rm d}t_{r}}{{\rm d}x}\frac{{\rm d}A_{\rm ce}}{{\rm d}t_{r}} \ .
\end{eqnarray}
where we introduced $A^{\rm ce}$ and $A^{\rm geo}$ the potentials related respectively to the charge excess and the geomagnetic emissions and $t_{r}$, the retarded time. Then, from the derivatives of $t_{r}$~\cite{2008APh....29...94S} the charge excess to geomagnetic ratio yields
\begin{eqnarray}
    a\equiv\sin{\alpha}\frac{E_{\rm ce}}{E_{\rm geo}}& \approx& \sin{\alpha}\frac{C_{x}}{\langle v_{d} \rangle}\frac{{\rm d}t_{r}}{{\rm d}x} \left(\frac{{\rm d}t_{r}}{{\rm d}t}\right)^{-1}\\
    &\approx& \sin{\alpha}\frac{C_{x}}{\langle v_{d} \rangle}\frac{x}{R} \\
    &\approx& \sin{\alpha}\frac{C_{x}(\rho)}{\langle v_{d}(\rho) \rangle}\sin{\omega} \ , \label{eq:scaling_ce_geo}
\end{eqnarray}
with $C_{x}$ being the fraction of particles  that  contributes  to the charge excess with respect to the total number of positrons and electrons, $R$ the distance between the source and the observer, $\langle v_{d} \rangle$ the mean drift velocity of the electrons in the air-shower plasma (in units of $c$) and $\alpha$ the geomagnetic angle. The dependency with $\omega$ indicates that for the radio emission of a given air-shower, the charge excess to geomagnetic ratio will not be spatially uniform. Usually for $\omega > 4\degree$ the signal amplitude is too low to be measured, then restricting to $\omega$ values below 4$\degree$ we can get the approximation $\sin{\omega} \approx \omega$ confirming the linear increase of the charge excess to geomagnetic ratio with $\omega$ and hence with distance to the shower core. \\

Another interesting feature in Fig.~\ref{fig:ce_geo_w_starshape} is the presence of a peak in the ratio around $\omega$ = $1$\degree. This peak was also noticed by~\cite{2010APh....34..267D, 2016APh....74...79K} but its origin remains debated \cite{2016APh....74...79K,2013APh....45...23D}. Cerenkov-like effects could be linked to this peak.
For example, if the charge excess emission peaks deeper in the atmosphere than the geomagnetic emission, its  Cerenkov cone is expected to be closer to the shower axis than the Cerenkov cone related to the geomagnetic emission due to density effects. Consequently, we expect an increase in the ratio for an $\omega$ angle corresponding to the maximum of emission of the charge excess and then a decrease as we reach the the maximum related to the geomagnetic contribution, resulting in a peak for the ratio. \\

In Eq.~(\ref{eq:scaling_ce_geo}), we also highlighted the dependency of the charge excess to geomagnetic ratio with the density of the medium $\rho$ as it was also well highlighted and parametrized from CoREAS simulations in~\cite{Glaser_2017}. This dependency indicates that the charge excess to geomagnetic ratio should also vary with the geometry of the showers as inclined showers for example are expected to develop in thinner atmosphere than vertical air-showers. Particularly, the mean drift velocity $\langle v_{d} \rangle$ is expected to decrease with increasing density as the mean free path of electrons and positrons before collision with air atoms will be shorter. At the same time, for a denser medium we expect larger values of $C_{x}$ as there will be more air-atoms colliding with the air-shower, resulting in a larger number of electrons in the shower front. This feature will specifically be detailed in the next section.

\subsection{Dependency of the $a$ ratio on primary energy and zenith inclination}\label{section:ce_geo_dependencies}

\begin{figure*}[tb]
\centering 
\includegraphics[width=0.95\columnwidth]{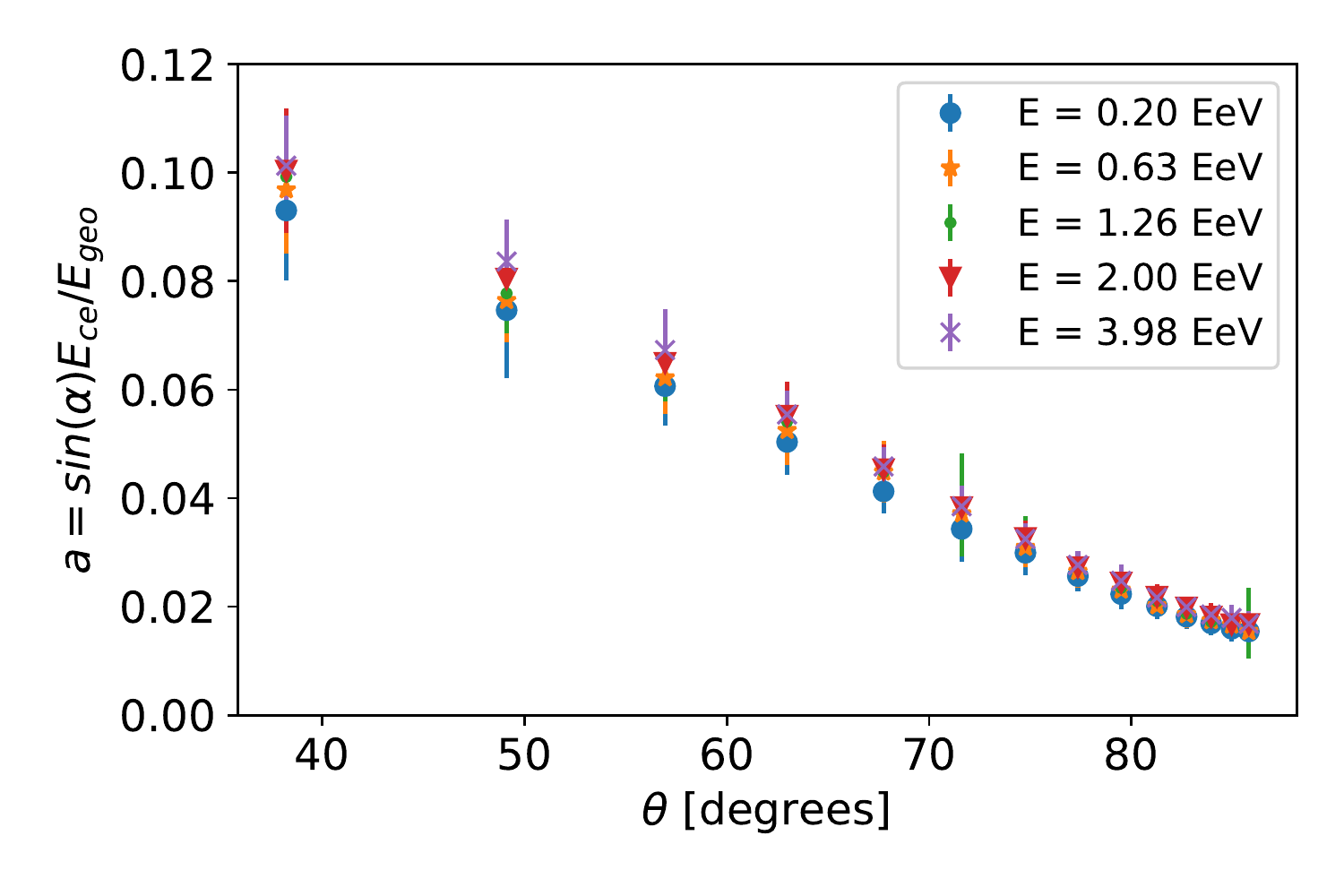}
\includegraphics[width=0.95\columnwidth]{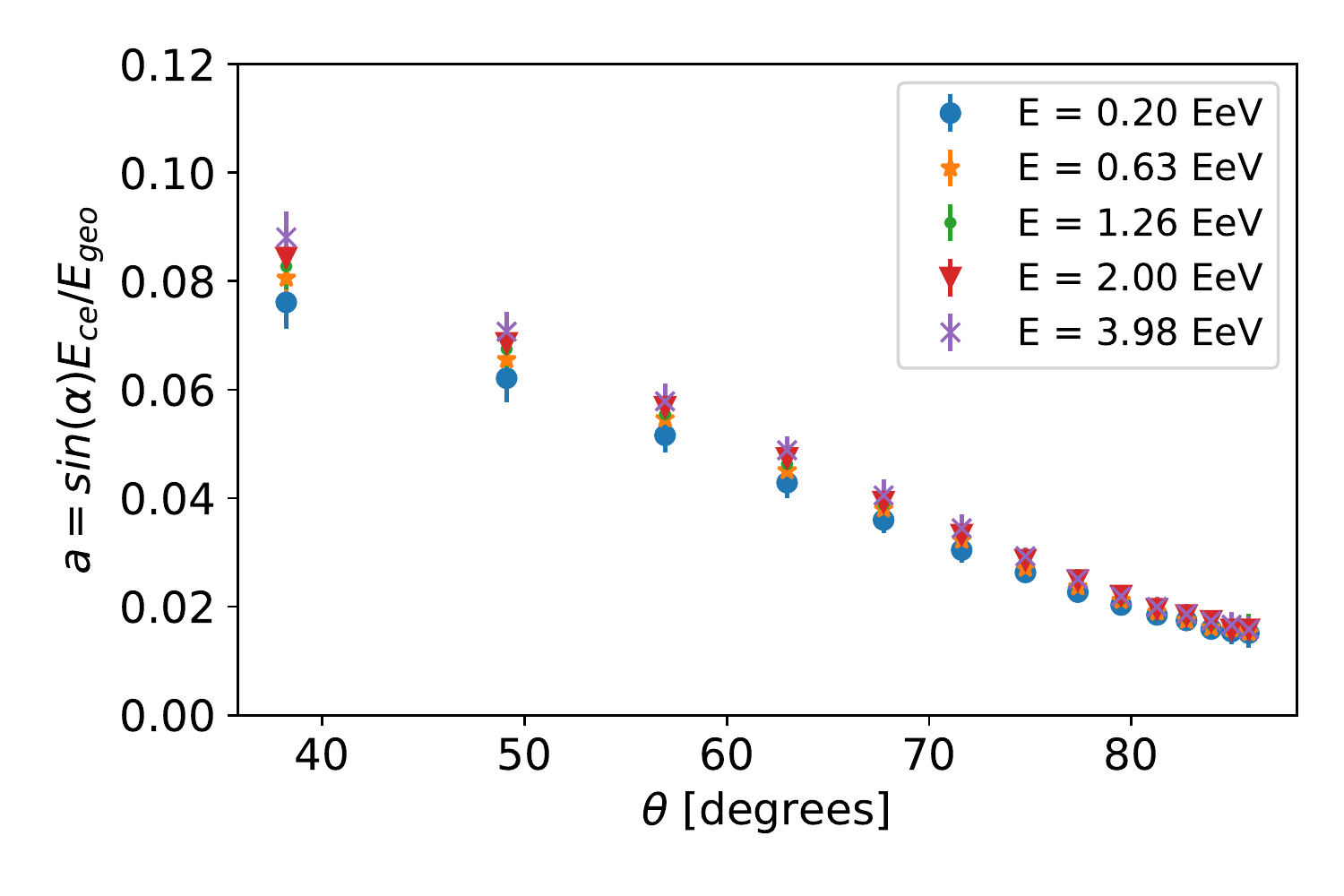}
\caption{Ratio between the charge excess to geomagnetic radiated energies corrected from the geomagnetic angle for raw traces of proton- ({\it left}) and iron- ({\it right})  induced showers as a function of the zenith angle and the energy of the primary particle. From a set of $\sim 5\,000$ simulations, we average the $a$-ratio of simulations having the same couple of zenith angle and energy, the error bars correspond to the RMS of each distribution. The radiated energy of both emission are derived by integrating the fluence of antennas along the  ${\bf v \times (v\times B)}$ axis and assuming a radial symmetry of the radio signal.\label{fig:ce_geo_vs_zenith_nop_proton} }
\end{figure*}

The dependency of the $a$-ratio with zenith angles and energy was discussed in~\cite{Glaser_2017}. For the purpose of self-consistency, we derive these dependencies from our set of simulations.  We first consider a set of $\sim5\,000$ simulations of proton- or iron-induced showers with various energies and directions (14 bins of energy and 5 bins of zenith angles corresponding to 75 simulations for each couple of energy and zenith, see Table~\ref{Table:shower_parameters_cr_stshp}) and apply the method presented in Section~\ref{section:polarisation_reconstruction} to reconstruct the charge excess and geomagnetic energy. We do so by integrating the fluence of antennas along the  { $\mathbf{v \times (v\times B)}$} axis, where both emissions are uncorrelated following \cite{Glaser_2016} and assuming a radial symmetry of the radio signal in the shower plane. This was first done for raw, unfiltered traces, i.e., traces before any processing detailed in Section~\ref{section:zhaires}. In Fig.~\ref{fig:ce_geo_vs_zenith_nop_proton},  we represent the ratio between the charge excess and geomagnetic energies as a function of the primary energy and zenith angle for simulations with proton and iron primaries.

\begin{table*}[tb]
\begin{center}
\begin{tabular}{lllllllll}
\hline
\begin{tabular}[c]{@{}l@{}}Energy\\ (EeV)\end{tabular}  & 0.20 & 0.63 & 1.26 & 2.00 & 3.98 &      &      &      \\
\hline
\begin{tabular}[c]{@{}l@{}}Zenith $\theta$\\ (Deg.)\end{tabular} & 38.2 & 49.1 & 57.0 & 63.0 & 67.8 & 71.6 & 74.8 & 77.4 \\
                                                         & 79.5 & 81.3 & 82.7 & 83.9 & 85.0 & 85.8 & 86.5 & 87.1 \\
                                                        \hline
\begin{tabular}[c]{@{}l@{}}Azimuth $\phi$\\ (Deg.)\end{tabular} & 0    & 90   & 180  &      &      &      &      &   \\
\hline
\end{tabular}
\caption{Lists of the primary particle energies $E$, zenith angles $\theta$ and azimuth angles $\phi$ of the arrival direction considered in Section~\ref{section:ce_geo_dependencies}.}\label{Table:shower_parameters_cr_stshp}
\end{center}
\end{table*}

\begin{table*}[tb]
\begin{tabular}{lllllllllllll}
\hline
\begin{tabular}[c]{@{}l@{}}Energy E\\   (EeV)\end{tabular} &
  0.020 &
  0.025 &
  0.032 &
  0.040 &
  0.050 &
  0.063 &
  0.079 &
  0.100 &
  0.126 &
  0.158 &
  0.200 &
  0.251 \\
 &
  0.316 &
  0.398 &
  0.501 &
  0.631 &
  0.794 &
  1.000 &
  1.259 &
  1.585 &
  1.995 &
  2.511 &
  3.162 &
  3.981 \\
  \hline
\begin{tabular}[c]{@{}l@{}}Zenith $\theta$\\   (Deg.)\end{tabular} &
  81.26 &
  82.72 &
  83.94 &
  84.95 &
  85.80 &
  86.50 &
  87.08 &
   &
   &
   &
   &
   \\
   \hline
\begin{tabular}[c]{@{}l@{}}Azimuth $\phi$\\    (Deg.)\end{tabular} &
  0 &
  90 &
  180 &
   &
   &
   &
   &
   &
   &
   &
   &
   \\
   \hline
\end{tabular}
\caption{Lists of the primary particle energies $E$, zenith angles $\theta$ and azimuth angles $\phi$ of the arrival direction considered in Section~\ref{section:trigger_thresholds}.}\label{Table:shower_parameters_cr_stshp_II}
\end{table*}

\begin{table*}[tb]
\begin{center}
\begin{tabular}{llllllllll}
\hline
\begin{tabular}[c]{@{}l@{}}Energy\\ (EeV)\end{tabular}   & 0.63  & 0.79  & 1.00  & 1.25  & 1.58 & 1.99 & 2.51  & 3.16 & 3.98 \\
\hline
\begin{tabular}[c]{@{}l@{}}Zenith $\theta$\\ (Deg.)\end{tabular}  & 81.26 & 82.72 & 83.94 & 84.95 & 85.8 & 86.5 & 87.08 &      &      \\
\hline
\begin{tabular}[c]{@{}l@{}}Azimuth $\phi$\\ (Deg.)\end{tabular} & 0     & 90    & 180   & 270   &      &      &       &      &    \\
\hline
\end{tabular}
\caption{Lists of the primary particle energies $E$, zenith angles $\theta$ and azimuth angles $\phi$ of the arrival direction considered in Section~\ref{section:HS1_layout}.}\label{Table:shower_parameters_cr_HS1}
\end{center}
\end{table*}

From these figures, it appears clearly that there is a mild correlation with the energy of the primary particle and a strong one with zenith angle. This can be explained by density arguments:	

Regarding the correlation with energy, it is expected that the $X_{\rm max}$ position of the air-shower should scale logarithmically with the energy of the primary~\cite{Heitler:1936jqw}. This implies that the density at $X_{\rm max}$ should also increase with increasing energy. Hence considering that most of the radio emission occurs at the maximum air-shower development depth $X_{\rm max}$, it appears in Eq.~\eqref{eq:scaling_ce_geo} that a higher density at $X_{\rm max}$ will result in a stronger charge excess, due to an enhanced accumulation of negative charges (increase of $C_{x}$) and a weaker geomagnetic emission (decrease of $\langle v_{d} \rangle$).

Regarding the dependency with the zenith angle, the $X_{\rm max}$ position is shallower for inclined showers, hence the charge excess to geomagnetic ratio lower, following Eq.~\eqref{eq:scaling_ce_geo}. 

\begin{figure}[tb]
\centering 
\includegraphics[width=0.95\columnwidth]{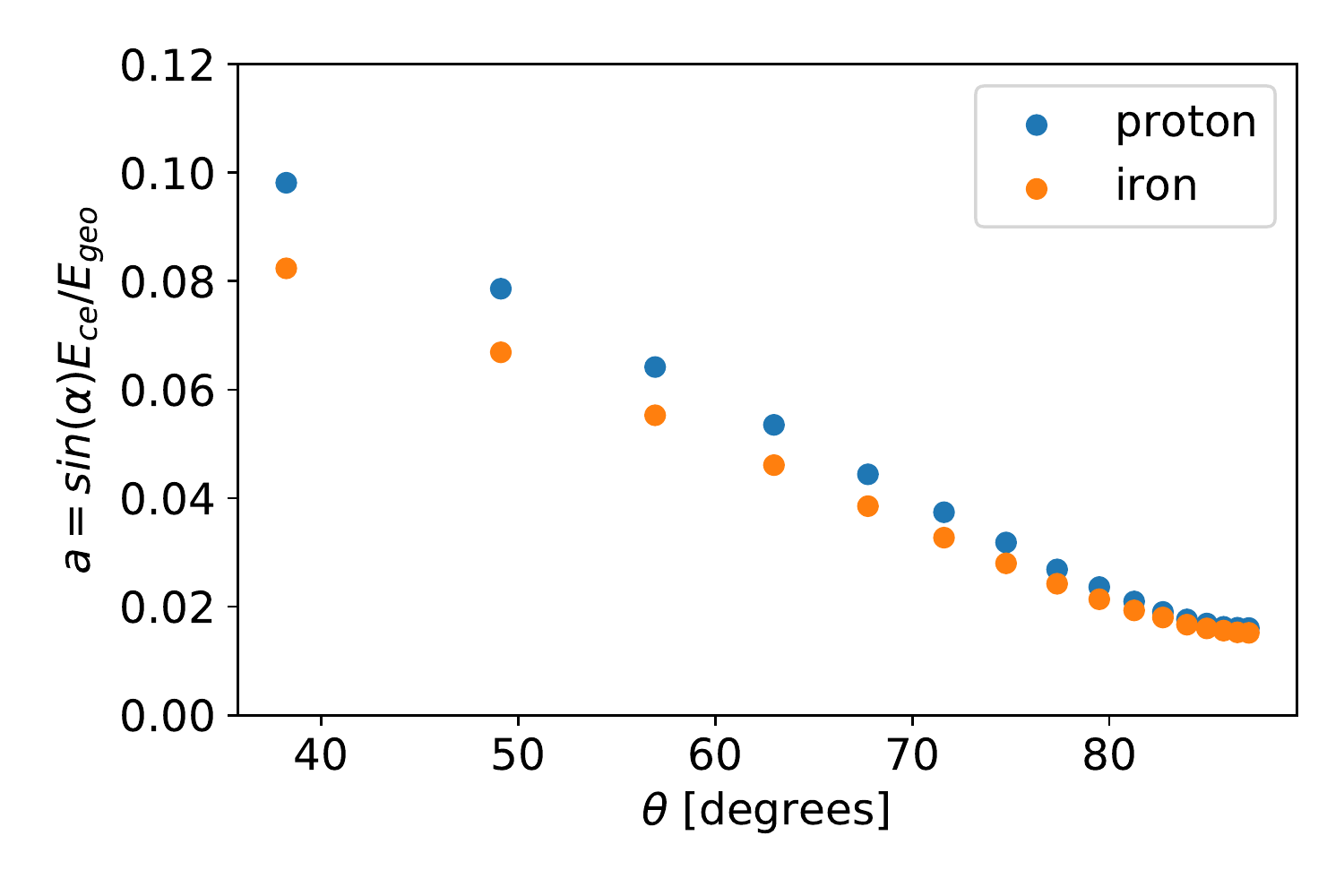}
\caption{Charge excess to geomagnetic ratio on average for simulations with proton or iron primaries and various energies}\label{fig:ce_geo_iron_vs_proton_all_zenith}
\end{figure}

Moreover, the shower maximum $X_{\rm max}$ should be shallower for iron than proton nuclei as the higher mass number increases the interaction probability~\cite{Huege:2016veh}. This implies that the charge excess to geomagnetic ratio is expected to be lower for iron-induced showers. This is confirmed with simulations as illustrated in Fig.~\ref{fig:ce_geo_vs_zenith_nop_proton}. To highlight this point better, in Fig.~\ref{fig:ce_geo_iron_vs_proton_all_zenith}, we averaged values over the energy bins.

\begin{figure}[tb]
\centering 
\includegraphics[width=0.95\columnwidth]{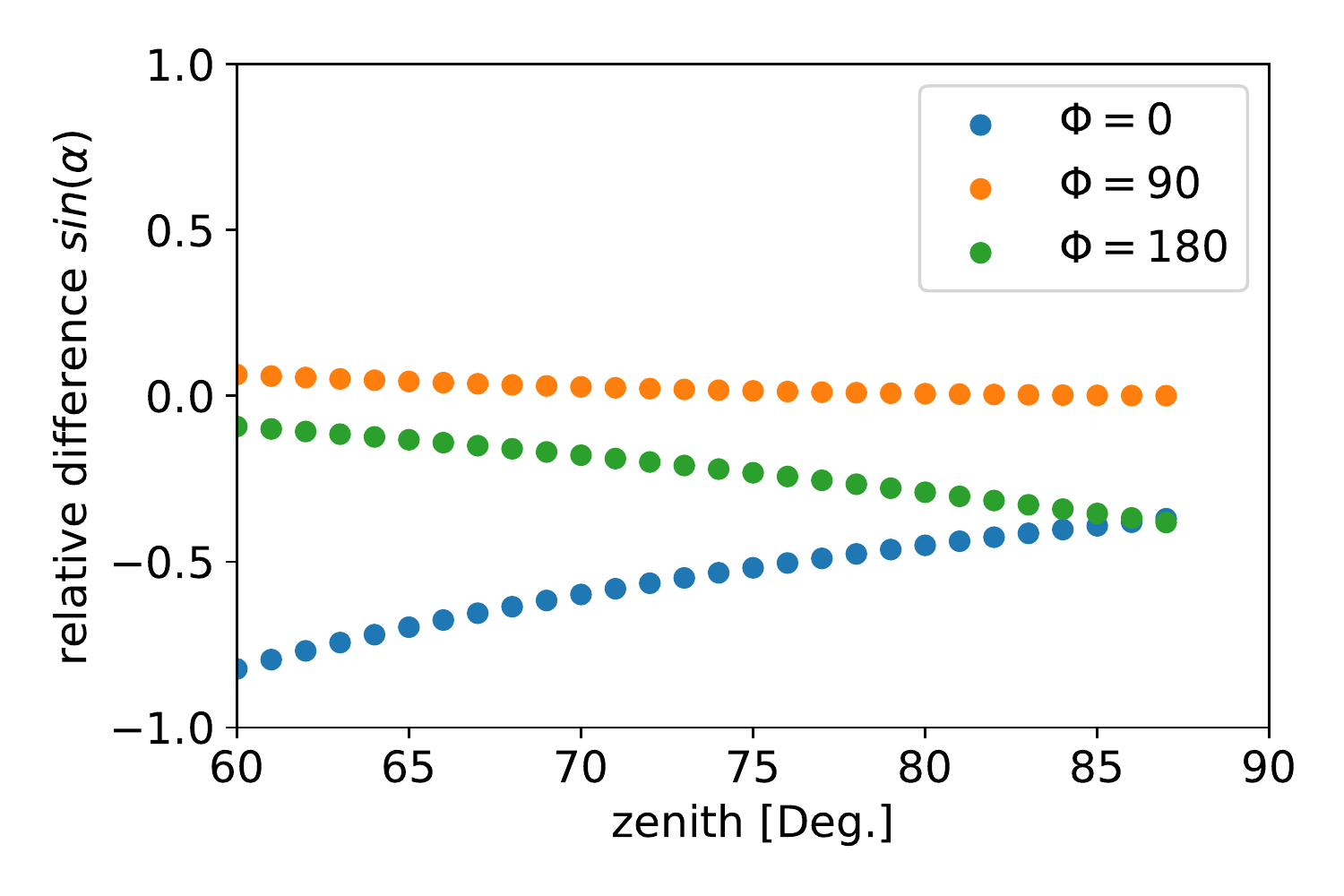}
\caption{ Relative difference of $\sin{\alpha}$ ($\alpha$ being the geomagnetic angle): $(\sin{\alpha_{\rm Dunhuang} - \sin{\alpha_{\rm Auger})/\sin{\alpha_{\rm Dunhuang}}}}$ as a function of the shower azimuth and zenith angle between the Dunhuang site with magnetic field inclination $I_{B} = 60.79$\degree and  the Auger site with $I_{B} = 35.2$\degree. $\Phi = 0$ corresponds to a shower towards North and the azimuth angle is counted positively towards Est.}\label{fig:geomagnetic_angle}
\end{figure}

To measure the impact of the geomagnetic angle on our results, we represent in Fig.~\ref{fig:geomagnetic_angle} the relative deviation between $\sin{\alpha}$ at the GRAND site with magnetic inclination $I_{B} = 60.79$\degree (left) and at the Auger site with magnetic inclination $I_{B} = 35.2$\degree (right). We find that although the Auger site is a location with particularly high Askaryan contribution to the radio-emission, the relative deviation of $\sin{\alpha}$ between the GRAND and the Auger site is only up to few tens of percent for very inclined showers. Significant deviations are observed only for showers with azimuth angle towards North and with low zenith angle (close to 60\degree). Between different locations on Earth, there are also variations of the magnetic field strength that can be roughly up to a factor 2. This implies that for very inclined showers ($\theta> 80\degree$) the combined effect of the geomagnetic angle and the magnetic field strength can lead up to a factor 4 between the Askaryan ratio at a favorable site for the geomagnetic emission (such as the GRAND site) and a non favorable site (Auger site). Still, as we can see in Figure~\ref{fig:ce_geo_vs_zenith_nop_proton}, for the GRAND site at 80\degree, the Askaryan ratio is of $\sim 2\%$ which implies that even with an additional factor 4 we expect a dominant geomagnetic emission. This justifies the assumption of a radio-emission dominated by the geomagnetic mechanism for inclined showers, independently of the location on Earth.

\subsection{Effects of signal processing and trigger conditions}

We now perform a similar study for realistic time traces, i.e., applying to the traces the full treatment described in Section~\ref{section:zhaires} and then a trigger condition at the antenna level on the reconstructed amplitude $\sqrt{I}$ of either $E_{\rm trig, 3\sigma}=60\,\mu$V/m (3$\sigma$ trigger threshold, left panel) or $E_{\rm trig, 5\sigma}=100\,\mu$V/m (5$\sigma$ trigger threshold, right panel). In the following, we will always present results for 3 distinct cases: no processing (i.e. raw traces), 3$\sigma$ trigger threshold ($\sigma_{\rm noise} = 20 \mu$V/m, $E_{\rm trig, 3\sigma}= 60\mu$V/m + filtering and sampling) and 5$\sigma$ trigger threshold ($\sigma_{\rm noise} = 20 \mu$V/m, $E_{\rm trig, 5\sigma}= 100\mu$V/m + filtering and sampling). 

\begin{figure*}[tb]
\includegraphics[width=0.48\linewidth]{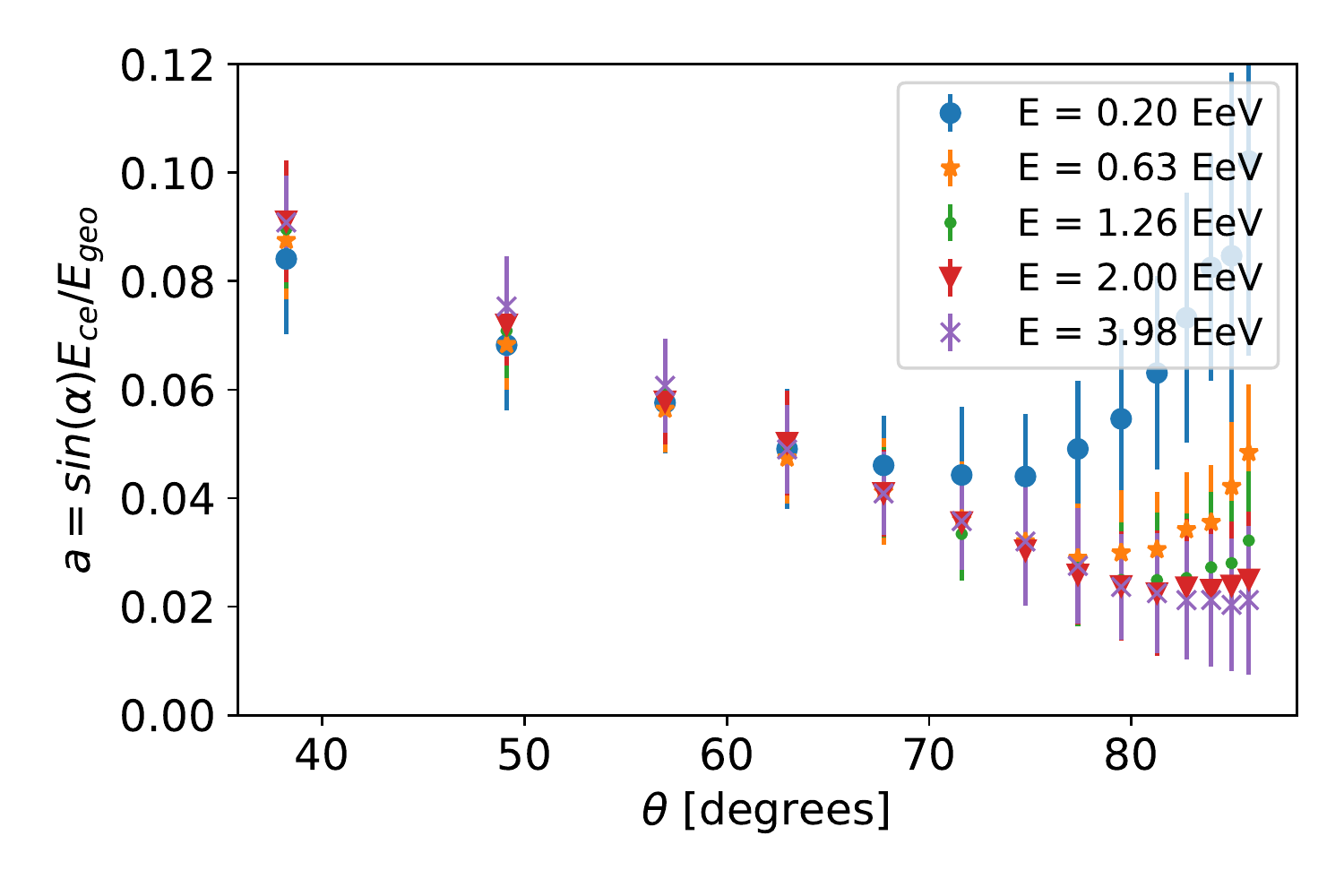}\hfill
\includegraphics[width=0.48\linewidth]{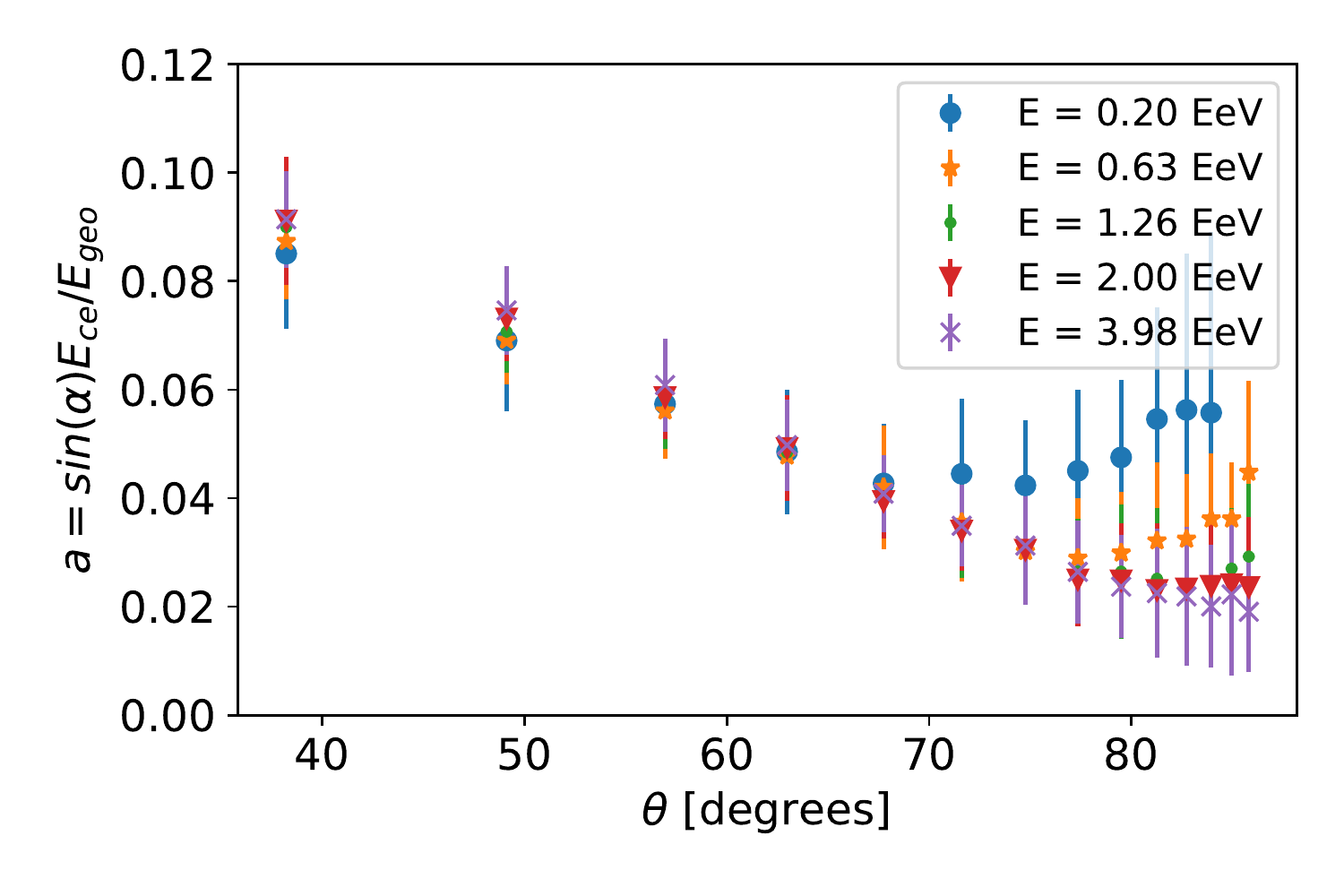}
\caption{Charge excess to geomagnetic ratio as a function of the energy and the zenith angle for simulations with a proton primary on a star-shape layout, considering traces with noise and a $3\sigma$ trigger threshold {\it (left)} or a $5\sigma$ trigger threshold {\it (right)}. Each data point corresponds to the average ratio for simulations with the same energy and zenith angle and the error bars to the RMS of each distribution.}\label{fig:ce_geo_vs_zenith_with_noise}
\end{figure*}

In Figure~\ref{fig:ce_geo_vs_zenith_with_noise}, we represent the $a$-ratio derived from Eq.~\ref{eq:norm_ce}, \ref{eq:norm_geo} and~\ref{eq:ce_geo} using this time-processed outputs. Note that this $a$-ratio does not directly correspond to the physical ratio between the charge excess and geomagnetic emission, as presented in the previous section, but rather to the expected measurement in an experimental case. We observe for showers with low inclination a similar behavior as in Fig.~\ref{fig:ce_geo_vs_zenith_nop_proton}. However, for most inclined showers ($\theta >70 \degree$), we can also observe that above a certain zenith angle the ratio increases with inclination, in particular at low energy.  This can be understood considering that, the more inclined the shower, the more diluted the radio signal, because the shower develops over longer distances \cite{Huege:2016veh}. As a consequence, for the most inclined showers we will have an increasing number of antennas with low amplitude signals (mostly for low energy showers). This implies that for such showers the amplitude of the charge excess mechanism is comparable to the noise amplitude, inducing a bias in the $a$ ratio.
 
 This plot also allows us to infer an interesting feature: for cosmic-ray inclined air-showers (with $\theta > 65^\circ$), the charge excess to geomagnetic ratio drops below 5\% independently of the trigger threshold for energies above 0.2\,EeV. As the geomagnetic emission is along ${\bf v} \times {\bf B}$, it is perpendicular to the local direction of the magnetic field. Consequently, in the case of a very dominant geomagnetic emission, which is expected for inclined air-showers, the total electric field should also be in good approximation perpendicular to the direction of the magnetic field. This is a specific feature that could enable to discriminate between the radio signal from cosmic-ray air-showers and the background noise, providing an efficient way to perform autonomous radio detection. 
 
In the following sections we quantify more precisely this effect for inclined air-showers induced by cosmic-rays and establish a criterium for a trigger at the DAQ level.

\section{Shower identification method}\label{section:identification_method}

In the previous section, we highlighted that radio emission from inclined EAS is strongly polarised in a direction orthogonal to $\mathbf{B}$, a specific feature which could thus be used to allow for autonomous radio-detection, i.e. a detection of air-showers using the radio signal only for antennas with 3 polarisation channels. In this section, similarly to what was presented in~\cite{Aab_2016_2} we define the trigger thresholds required to perform such an identification, and apply our method to a realistic array layout.

\subsection{Threshold definitions for a trigger at the DAQ level}\label{section:trigger_thresholds}
\subsubsection{Projection of the electric field along $\mathbf{B}$}

In this section, we focus only on showers with a zenith angle between 80$\degree$ and 90$\degree$. We consider a set of  $\sim 11\,000$ simulations for proton and iron-induced showers, with primary particle energies, azimuth and zenith angles given in Table~\ref{Table:shower_parameters_cr_stshp_II}. 
We aim at establishing a criterium to trigger through an online treatment at the DAQ level with a strong background rejection. For this purpose, we focus on the fraction of the total electric field along the direction of the magnetic field, i.e., $E_{b}/E_{\rm tot}$, where $E_{\rm tot}$ is the total amplitude of the electric field, ${E_{\rm tot}} = \sqrt{E_{x}^{2} + E_{y}^{2} + E_{z}^{2}}$ and $E_{b}$ is the absolute value of the projection of the total electric field along the direction of {${\bf B}$}, {$E_{b} = |\mathbf{E}_{\rm tot} \cdot \mathbf{u}_{b}$|} with $\mathbf{u_{b}}$ the unit vector along the direction of the magnetic field. As the geomagnetic emission is perpendicular to ${\bf B}$, the only contribution to $E_{b}$ comes from the charge excess and for a dominant geomagnetic emission $E_{\rm tot}$ is similar to $E_{\rm geo}$. Therefore, the ratio $E_{b}/E_{\rm tot}$ should be quite similar to the $a$ ratio. However using this ratio has 2 main advantages:

\begin{itemize}
\item First, it can be computed at the DAQ level directly after a preliminary reconstruction of the direction of origin of the radiation. a necessary step to deconvolve the antenna response from the voltage signal and thus provide a fast estimate of the electric field amplitude, while for the $a$-ratio, calculations have to be performed using the amplitude of the electric field in the shower plane, which requires information about the shower geometry. The reconstruction of the shower direction for very inclined air-showers has been studied in~\cite{2011} and shows that a plane angular reconstruction at the DAQ level of 1{\degree} should be achievable from the raw timing information. The impact of misreconstructed events in our background rejection efficiency is however not considered in this paper and should be quantified in further studies.

\item Second, as $E_{b}$ corresponds to a projection of the charge excess, we have $E_{b}$<$E_{\rm ce}$ and since $E_{\rm geo}<E_{\rm tot}$, $E_{b}/E_{\rm tot}$ allows to put even more restrictive constraints on the trigger condition.
\end{itemize}

\begin{figure*}[tb]
\includegraphics[width=0.50\linewidth]{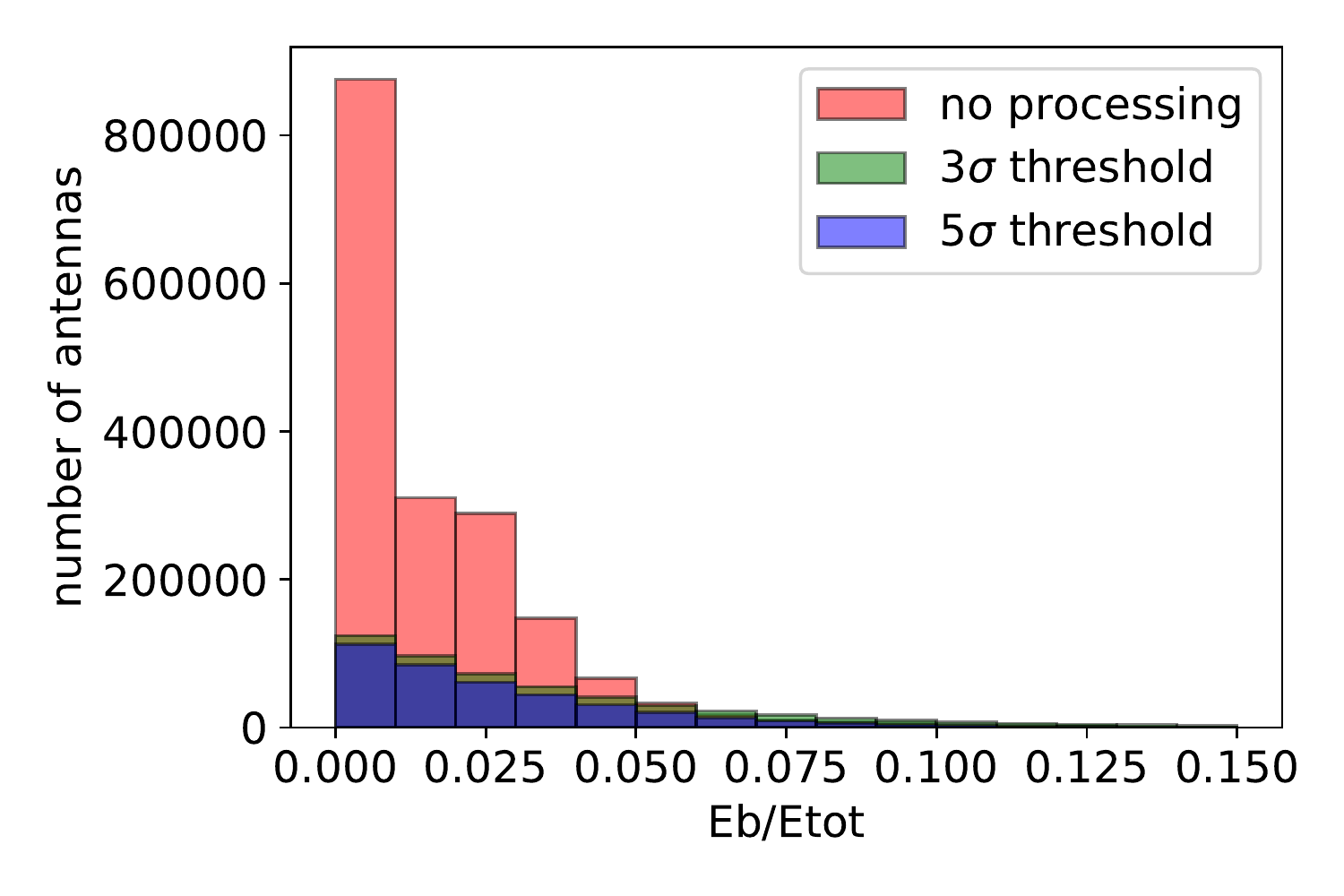}\hfill
\includegraphics[width=0.50\linewidth]{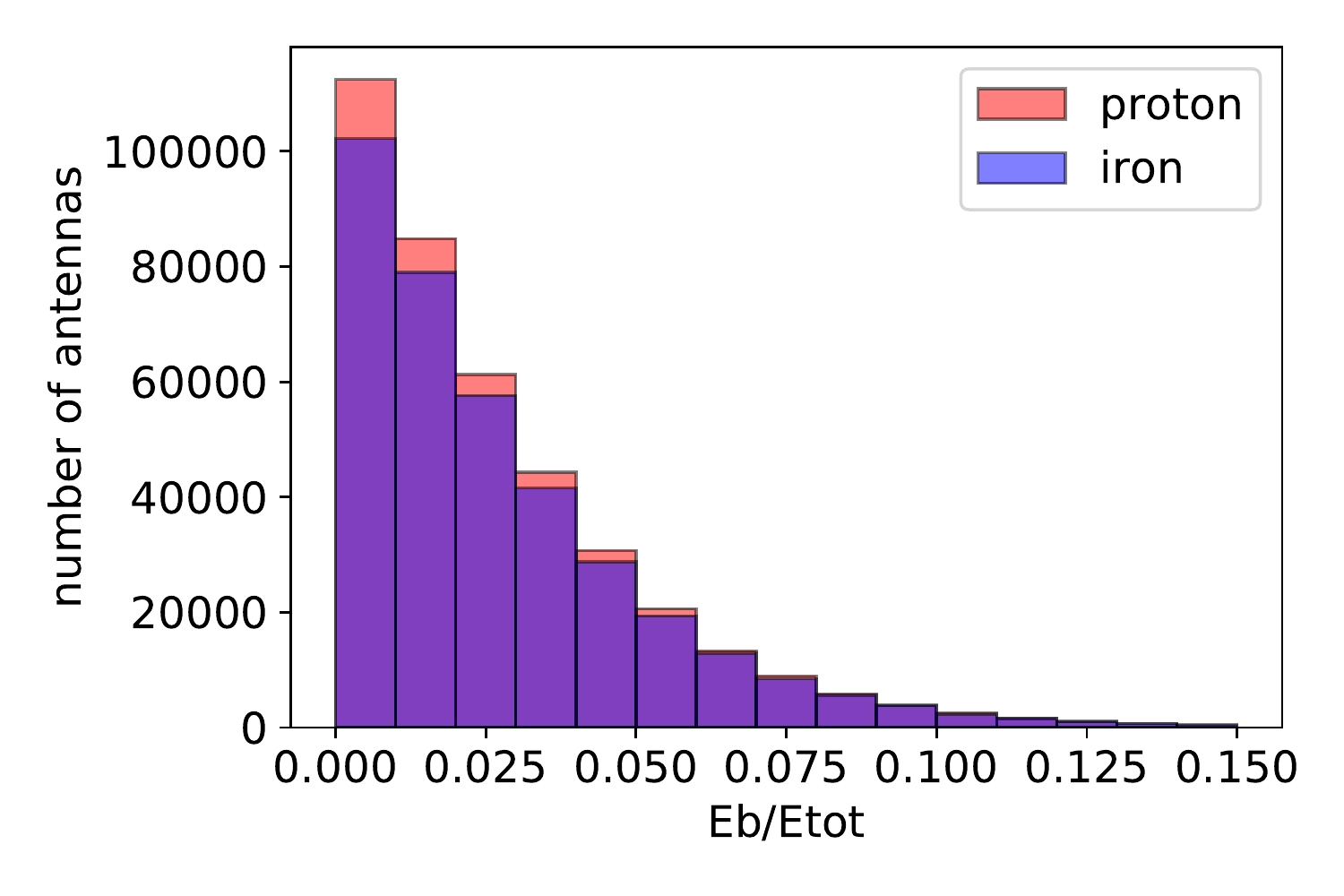}
\caption{ Histogram of the $E_{b}/E_{\rm tot}$ ratio measured at each antenna for a set of about $11\,000$ simulations for ({\it left}) proton induced showers with different trigger thresholds and ({\it right}) proton or iron induced showers and a $5\sigma$ trigger threshold.}\label{fig:Eb_ratio_antennas_iron_vs_proton}
\end{figure*}

In Fig.~\ref{fig:Eb_ratio_antennas_iron_vs_proton}, we represent histograms of the $E_{b}/E_{\rm tot}$ ratio at the antenna level for our set of $11\,000$ proton and iron simulations in the 3 following cases: no processing, 3$\sigma$ trigger threshold and 5$\sigma$ trigger threshold (left panel) and comparing proton and iron primaries at the 5$\sigma$ trigger level (right panel). Note that for histograms at the $3\sigma$ and $5\sigma$ trigger thresholds, the effective number of antennas is reduced as some are cut with the trigger condition. From the plot on the left panel we attest, as could already be inferred in Fig.~\ref{fig:ce_geo_vs_zenith_with_noise}, that the ratio is distributed towards higher values in the 3$\sigma$ and 5$\sigma$ case than in the no processing case because of the noise added. As expected, we also observe a slightly larger number of antennas with a $3\sigma$ threshold than a $5\sigma$ threshold, as we considered a more aggressive trigger condition. Considering now the right panel of  Fig.~\ref{fig:Eb_ratio_antennas_iron_vs_proton}, very similar results for proton and iron primaries can be found, in agreement with Fig.~\ref{fig:ce_geo_iron_vs_proton_all_zenith}.  	
The most interesting feature is that independently of the primary or the trigger threshold considered, the expected $E_{b}/E_{\rm tot}$ ratio drops to very low values for the large majority of antennas: it is below 7\% for 86\% (93\%) of antennas at the $3\sigma$ ($5\sigma$) trigger threshold. This is a strong signature for air-showers which can be used at the DAQ level.

\begin{figure*}[tb]
\includegraphics[width=0.50\linewidth]{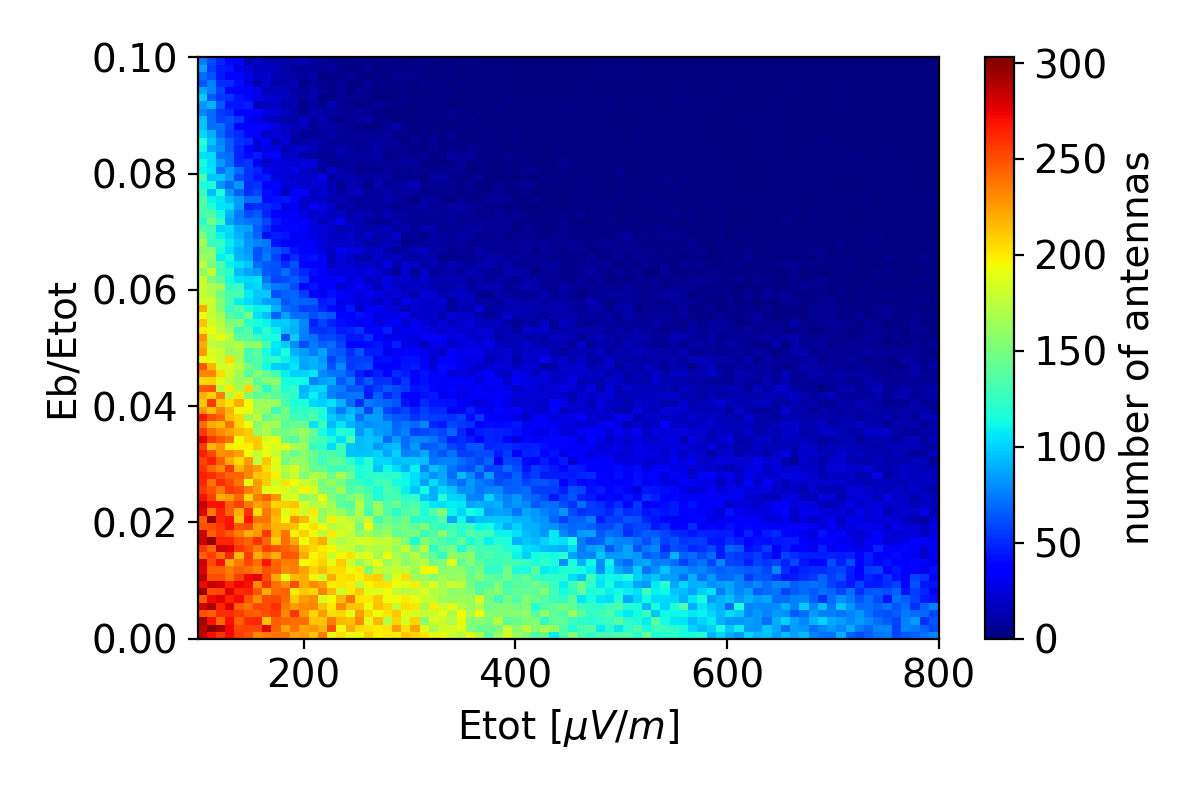}\hfill
\includegraphics[width=0.50\linewidth]{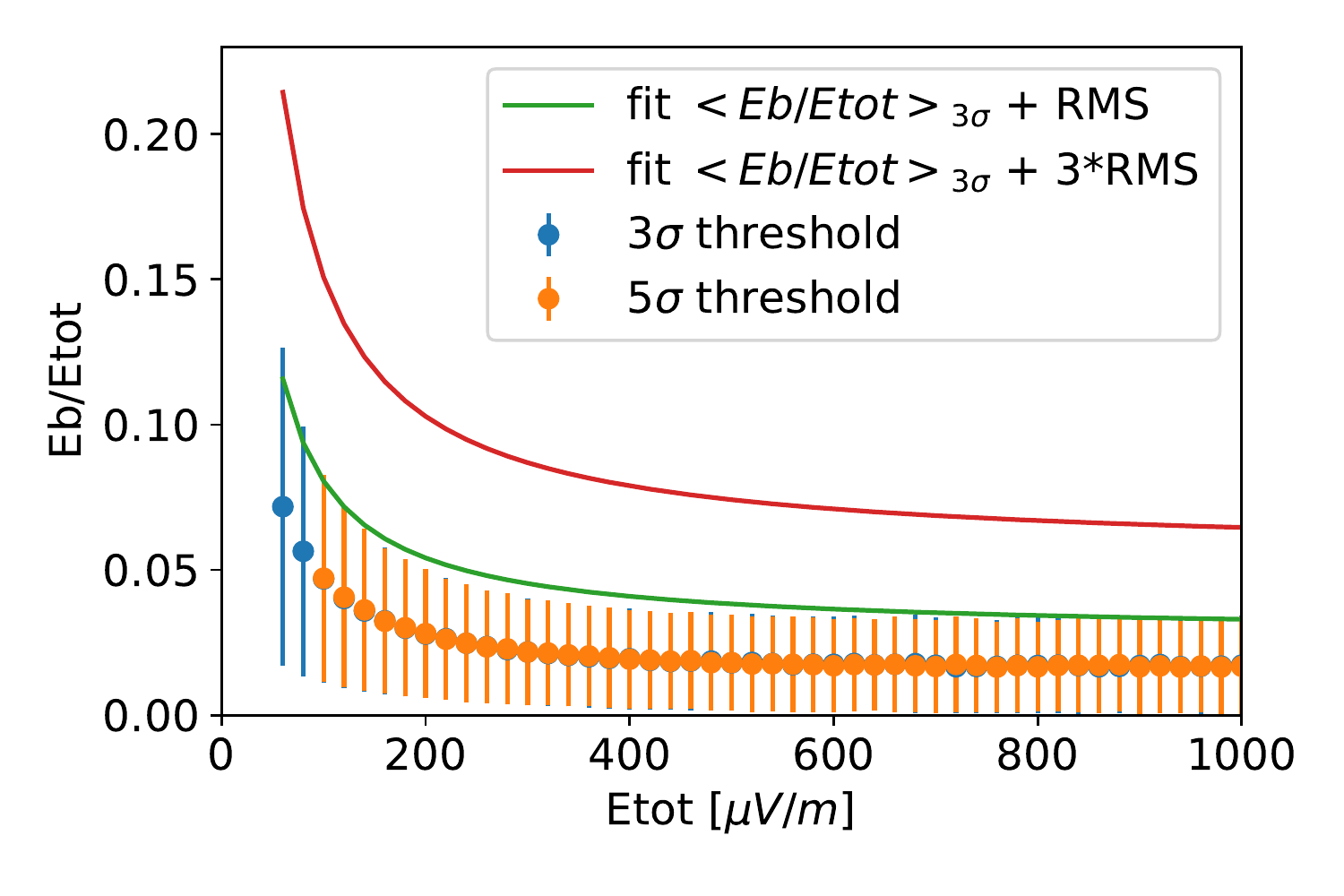}
\caption{({\it Left}) $E_{b}/E_{\rm tot}$ ratio as function of the total amplitude $E_{\rm tot}$ measured by each antenna on a star-shape layout for a set of $\approx \, 11\,000$ simulations on a star-shape layout considering a $5\sigma$ trigger threshold. ({\it Right}) $E_{b}/E_{\rm tot}$ ratio as function of the total amplitude measured by each antenna and averaged for different bins of the total amplitude, considering either a $3\sigma$ or a $5\sigma$ trigger threshold.}\label{fig:Eb_ratio_Etot}
\end{figure*}

In Fig.~\ref{fig:Eb_ratio_Etot}, we represent the distribution of the $E_{b}/E_{\rm tot}$ ratio as a function of the amplitude $E_{\rm tot}$ on a 2D histogram for the set of $11\,000$ simulations, at the  $5\sigma$ trigger threshold (left panel). On the right panel, we average the values of the ratio for different bins of the amplitude with either a 3$\sigma$ or 5$\sigma$ trigger threshold (right panel).  
It appears that the highest values of the $E_{b}/E_{\rm tot}$ ratio are related to antennas with low amplitude. This can be understood as for these antennas the charge excess amplitude is subdominant comparatively to the noise so that our measurement of the $a$ ratio ratio becomes closer to the noise/geomagnetic ratio. Hence, to strengthen our trigger criteria we could define a trigger condition that depends on the total amplitude measured by each antenna. We could for example define a trigger condition $E_{b}/E_{\rm tot}< f(E_{\rm tot})$ with $f(E_{\rm tot})$ obtained by fitting the points $\langle E_{b}/E_{\rm tot} \rangle + n\sigma_{E_{b}/E_{\rm tot}}$, $n$ being an integer and $\sigma$ corresponding here to the RMS of the distribution.

\begin{figure}[tb]
\centering 
\includegraphics[width=0.95\columnwidth]{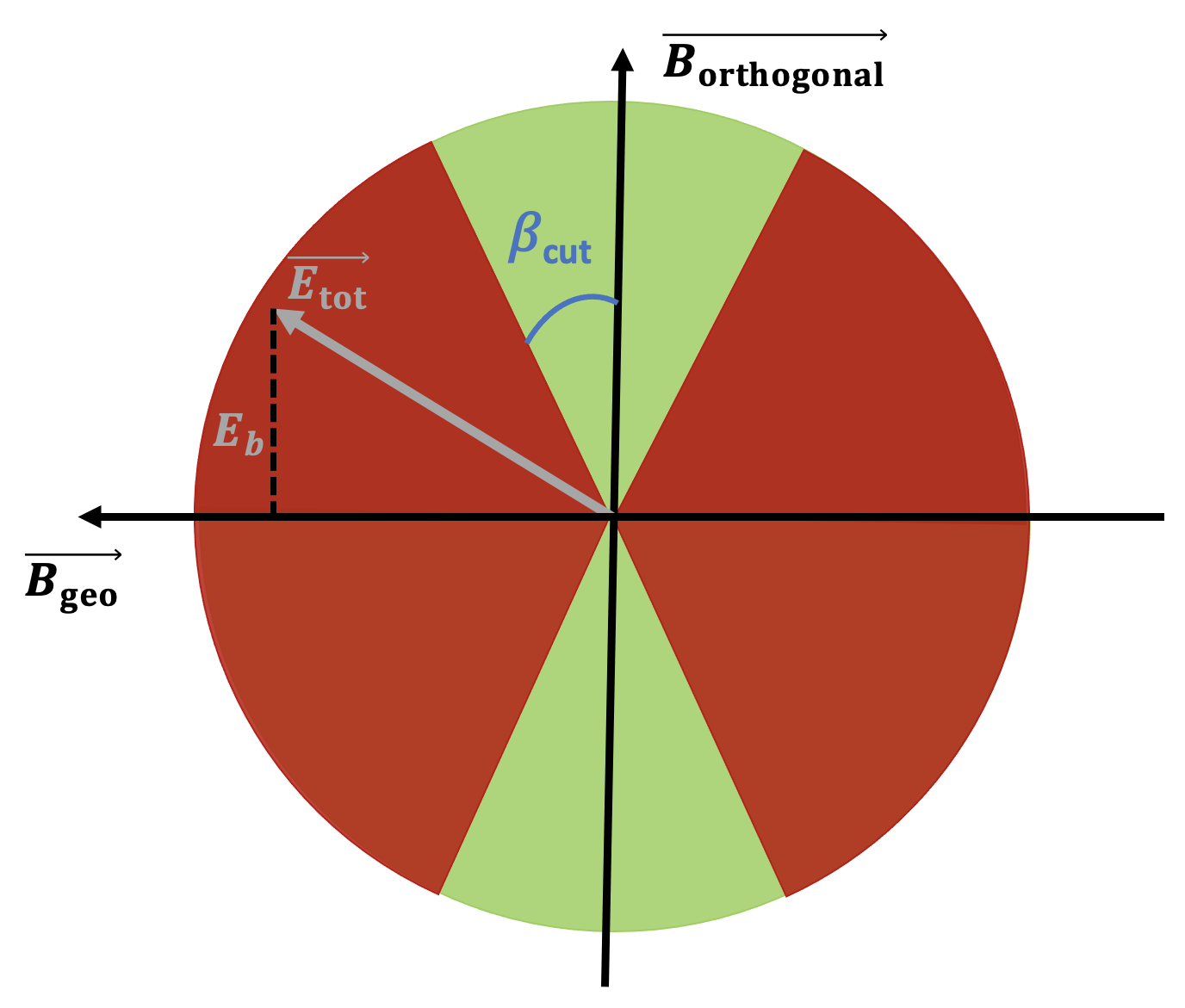}
\caption{Sketch of the electric field vector $E_{\rm tot}$ and its projection on the magnetic field of the Earth, $E_{b}$. With a fixed cut on the $E_{b}/E_{\rm tot}$ ratio, the field of view is restricted to the green area and all the red area corresponds to rejected events.} \label{fig:sketch_b_ratio}
\end{figure}

In Fig.~\ref{fig:Eb_ratio_Etot}, we fit $f(E_{\rm tot})$  with the following trigger function
\begin{equation}
    f(E_{\rm tot}) = a + \frac{b}{E_{\rm tot}} \label{eq:trigger_function}
\end{equation}
For $n=1$, we find $a = 0.1620 \pm 0.0005$ and $b = 6 \pm 0.1$ which results in a $83\%$ trigger efficiency and for $n =3$ we have $a = 0.0340 \pm 0.0009$ and $b = 11 \pm 0.3$ resulting in a $99\%$ trigger efficiency. As modeling the transient noise is challenging, it is not possible to quantify directly the fraction of background events that will be rejected. However, selecting an $E_{b}/E_{\rm tot}$ ratio  $\leq 7$\% restricts our field of view to only $7\%$ of the full solid angle as illustrated in Fig.~\ref{fig:sketch_b_ratio}).
Hence, assuming an isotropic polarisation for the noise orientation this cut should reduce the number of noise triggered events by a factor $\sim 15$ while more than 93\% of antennas pass this cut for cosmic-ray signals for a 5$\sigma$ threshold. This provides a strong noise rejection efficiency, even without knowing any of its features. To quantify the impact of the geographic location on the $E_{b}/E_{\rm tot}$ ratio, we can also compute our background rejection efficiency in the least favorable case. Following the text at the end of Section~\ref{section:ce_geo_dependencies}, for a non favorable site, the Askaryan ratio should scale by a factor 4 compared to the one computed at the GRAND site. Hence, the $E_{b}/E_{\rm tot}$ ratio should also roughly scale by a factor 4 which lead to a new value of $(E_{b}/E_{\rm tot})_{\rm non \,favorable} = 0.07\times 4 = 0.28$. With this new value we find a background rejection efficiency of $72\%$.

Yet the selection treatment presented here requires that the electric field amplitude is reconstructed by deconvolving the output voltage from the antenna response. As it is strongly anistropic, this treatment  requires that the direction of origin of the signal is known, which is therefore possible only at the central DAQ level, by combining the timing information of different triggered detection units. Interestingly enough, the raw voltage information available at the individual antenna level already provides some information on the nature of the signal and should allow for a first rough ---but efficient--- rejection of background signals.  
To illustrate this, we represent in Fig.~\ref{fig:Eb_Vb_comparison} the quantity $V_{b}/V_{\rm tot}$ (i.e., the voltage ratio computed in a same way as the $b$-ratio $E_{b}/E_{\rm tot}$ ) with a $5\sigma$ trigger threshold. Even though the orthogonality to the magnetic field vector is not as strong as for the electric field, we find that the voltage $b$ ratio is below $28\%$ for $90\%$ of antennas. Hence even though the limits on the voltage are not as stringent as the electric field, these results show that a trigger directly at the antenna level (i.e. at the first level trigger) could already be performed and reject 72\% of noise events.

{\subsubsection{Polarized intensity}

All the previous results concerning the identification of shower signatures were based on one unique observable, the $E_{b}/E_{\rm tot}$ ratio. The limits defined on this observable in the previous section seem already strong enough to perform an efficient background rejection. However, one may want to explore the possibility to highlight other criteria for an identification of shower signatures, as the combination of all the criteria would allow to put even more restrictive constraints on the trigger condition ensuring an even more robust autonomous detection. Here, we explore the degree of polarisation of the radio signal from air-showers. It has been well established that such signals should be highly polarised~\cite{Scholten:2017tcr}, thus we aim at quantifying this feature still in the perspective of a trigger at the DAQ level.

\begin{figure}[tb]
\centering 
\includegraphics[width=0.95\columnwidth]{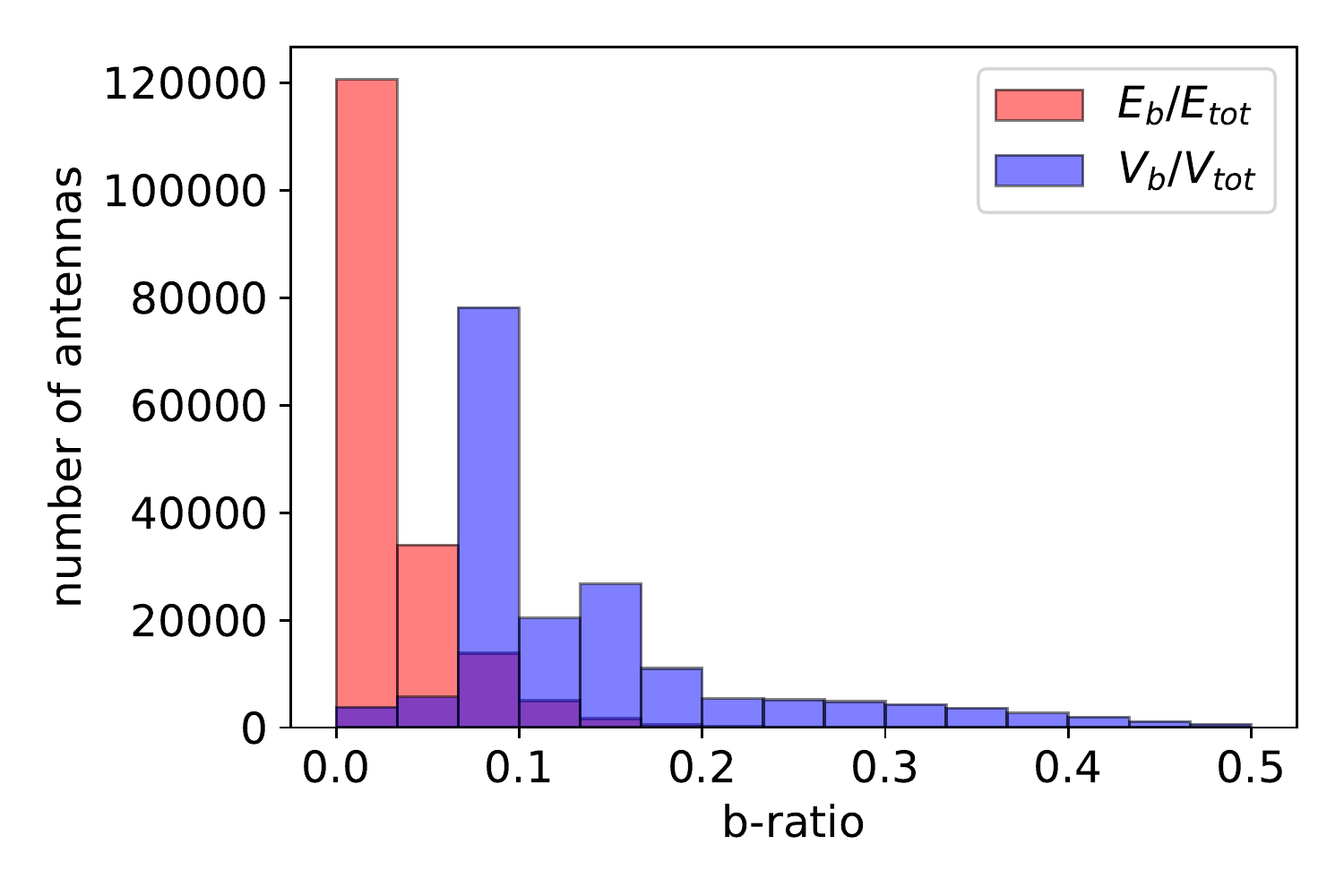}
\caption{ Comparison between of the ``$b$-ratio" for the electric field: $E_{b}/E_{\rm tot}$ and for the voltage: $V_{b}/V_{\rm tot}$.}\label{fig:Eb_Vb_comparison}
\end{figure}

The calculation of Stokes parameters as defined in Section~\ref{section:Stokes_parameters} requires only the components of the electric field. Then, instead of computing these parameters in the shower plane, we could also directly compute them at the DAQ level, using the 3 components in the geographic plane $E_{x}$, $E_{y}$, $E_{z}$. We can thus define the degree of polarisation of the signal as $I_{p}/I$, where $I$ corresponds to the Stokes parameter related to the total intensity of the signal and $I_{p} = \sqrt{Q^{2} + U^{2}+ V^{2}}$ corresponds to the polarised intensity. 

\begin{figure}[tb]
\includegraphics[width=0.95\columnwidth]{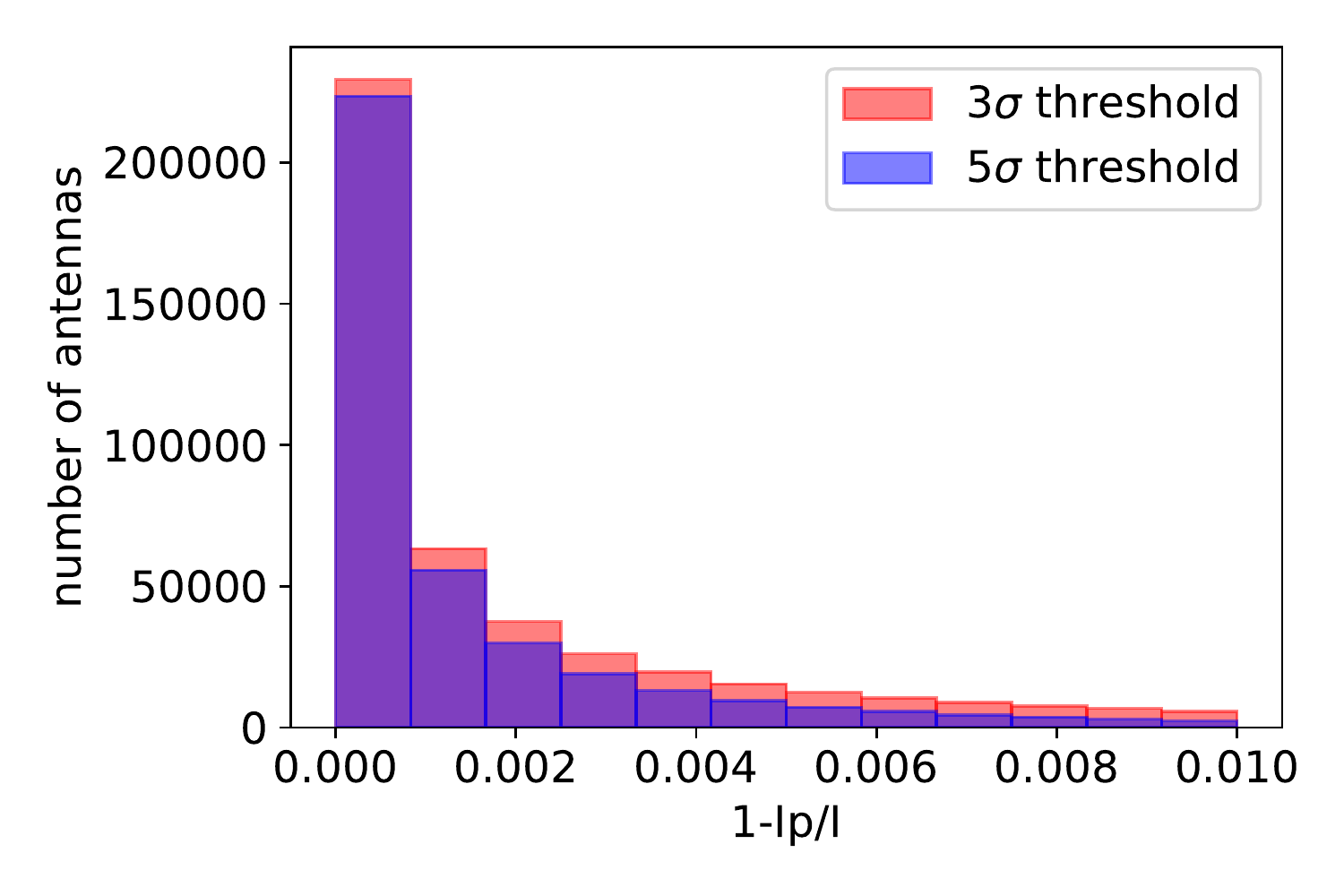}
\caption{ Histogram of $1 - I_{p}/I$ (where $I_{p}/I$ is the fraction of polarised intensity) measured at each antenna for proton induced showers on a star-shape layout at either a $3\sigma$ or a $5\sigma$ trigger threshold.}\label{fig:Ip_ratio_stshp}
\end{figure}

The distribution of $1-I_{p}/I$ is represented in Fig.~\ref{fig:Ip_ratio_stshp} for the star-shape simulation set}. We retrieve a highly polarised signal with more than $98\%$ of antennas having $1- I_{p}/I$ below $0.02$ at the $5\sigma$ trigger threshold.
This specific feature could also provide an efficient way to strengthen our criteria for autonomous radio detection as we could set very low limits on $1 - I_{p}/I$ without cutting the signal from any air-showers. It should however be noted that most of noises such as humans emissions are also expected to be highly polarised, hence this observable may not be as efficient as it seems. Still, it could be used as backup to reinforce the criterion defined in the previous section. Setting the limits of the $E_{b}/E_{\rm tot}$ ratio to 7\%, we already expect to cut $93\%$  of the noise, then among the 7\% remaining, the criterion over $I_{p}/I$ could help us evince all that are not strongly polarised, guaranteeing an even more efficient detection of air-showers.

\subsection{Application to a realistic layout}\label{section:HS1_layout}

\begin{figure}[tb]
\centering 
\includegraphics[width=0.99\columnwidth]{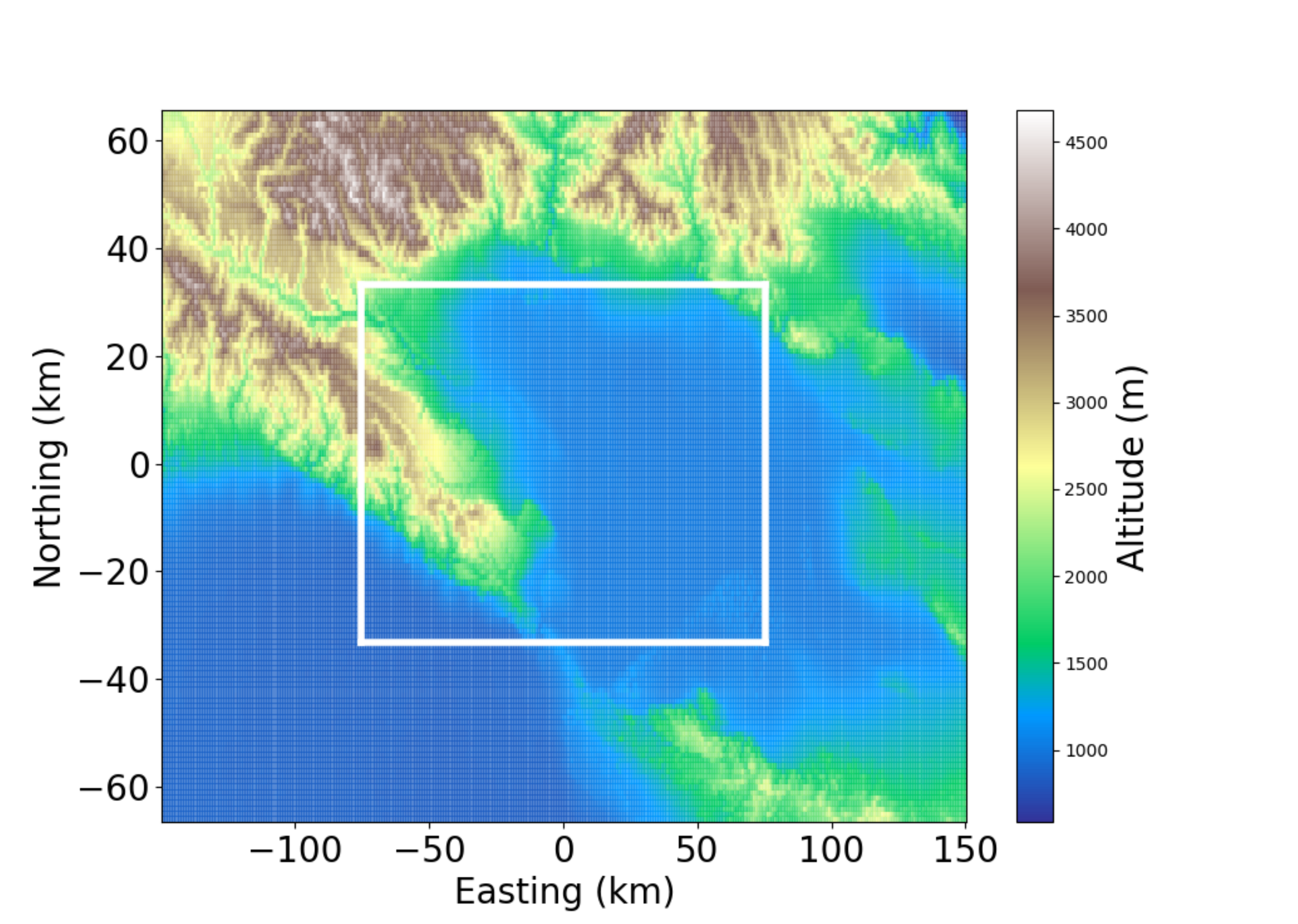}
\caption{Topography of the Hotspot 1 layout from \cite{2020SCPMA..6319501A}, corresponding to a giant rectangular array of $10\,000$ antennas over $10\,000$ square kilometers with simulated location in Ulastai (China).}\label{fig:HS1_layout}
\end{figure}

All the results in the previous section were presented for a star-shape layout which consists of a plane layout with a very dense array of antennas and for which the air-shower core is always centered on the layout center. In this section, we apply the same treatment to an experimental layout. The HotSpot 1 layout (hereafter HS1), is a rectangular layout made of $10\,000$ antennas~\cite{2020SCPMA..6319501A}, each with a spacing of 1 kilometer as illustrated in Fig.~\ref{fig:HS1_layout}. It corresponds to a real geographic location (centered at latitude $=42.1\degree$, longitude $=86.3\degree$) in the XinJiang province in China, used to evaluate the potential of GRAND for neutrino detection in \cite{2020SCPMA..6319501A}. 

Here, we consider a set of about $1\,000$ simulations of proton or iron induced showers corresponding to 9 energy bins, 7 zenith angles and 4 azimuth angles (see Table~\ref{Table:shower_parameters_cr_HS1}). For each simulation, the core position is generated randomly, resulting in a varying number of antennas inside the radio footprint.

\subsubsection{Projection of the electric field along \textbf{B}}

\begin{figure}[tb]
\includegraphics[width=0.95\columnwidth]{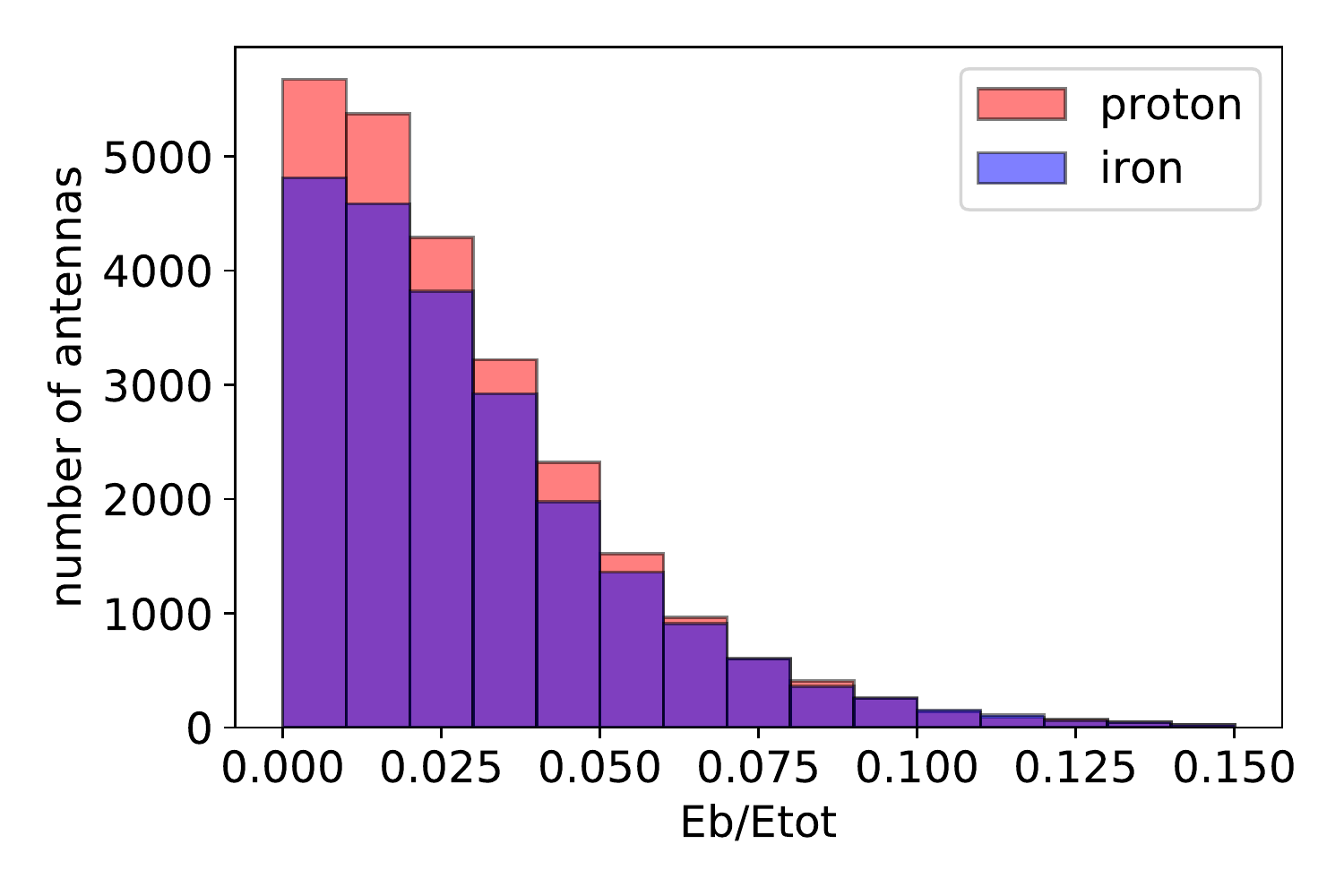}
\caption{ Histogram of the $E_{b}/E_{\rm tot}$ ratio measured at each antenna for proton and iron induced showers at the $5\sigma$ trigger threshold on the HS1 layout. }\label{fig:Eb_ratio_antennas_iron_vs_proton_HS1}
\end{figure}

In Fig.~\ref{fig:Eb_ratio_antennas_iron_vs_proton_HS1}, we present histograms of the $E_{b}/E_{\rm tot}$ ratio for proton and iron induced showers at the $5\sigma$ trigger threshold, similarly as in Fig.~\ref{fig:Eb_ratio_antennas_iron_vs_proton}. As previously observed, we retrieve for the  HS1 layout a strong similarity between proton and iron induced showers, suggesting that for inclined showers, the differences regarding the density at $X_{\rm max}$ for both of these primaries do not impact significantly the charge excess to geomagnetic ratio. Due to this high similarity, all the following results in this section will be presented for protons only. We also note that there is a larger number of antennas that trigger for proton than for iron simulations. This is related to the fact that the $X_{\rm max}$ position for iron-induced showers occurs higher in the atmosphere than for proton showers resulting in a larger footprint and a more diluted signal. Figure~\ref{fig:Eb_ratio_antennas_iron_vs_proton_HS1} also shows that even for an experimental layout such as HS1, we retrieve low values for the $E_{b}/E_{\rm tot}$ ratio, with more than 87\% of antennas with a ratio below 7\% at the $3\sigma$ trigger threshold level and about 93\% at the $5\sigma$ trigger threshold level.

\begin{figure}[tb]
\includegraphics[width=0.95\columnwidth]{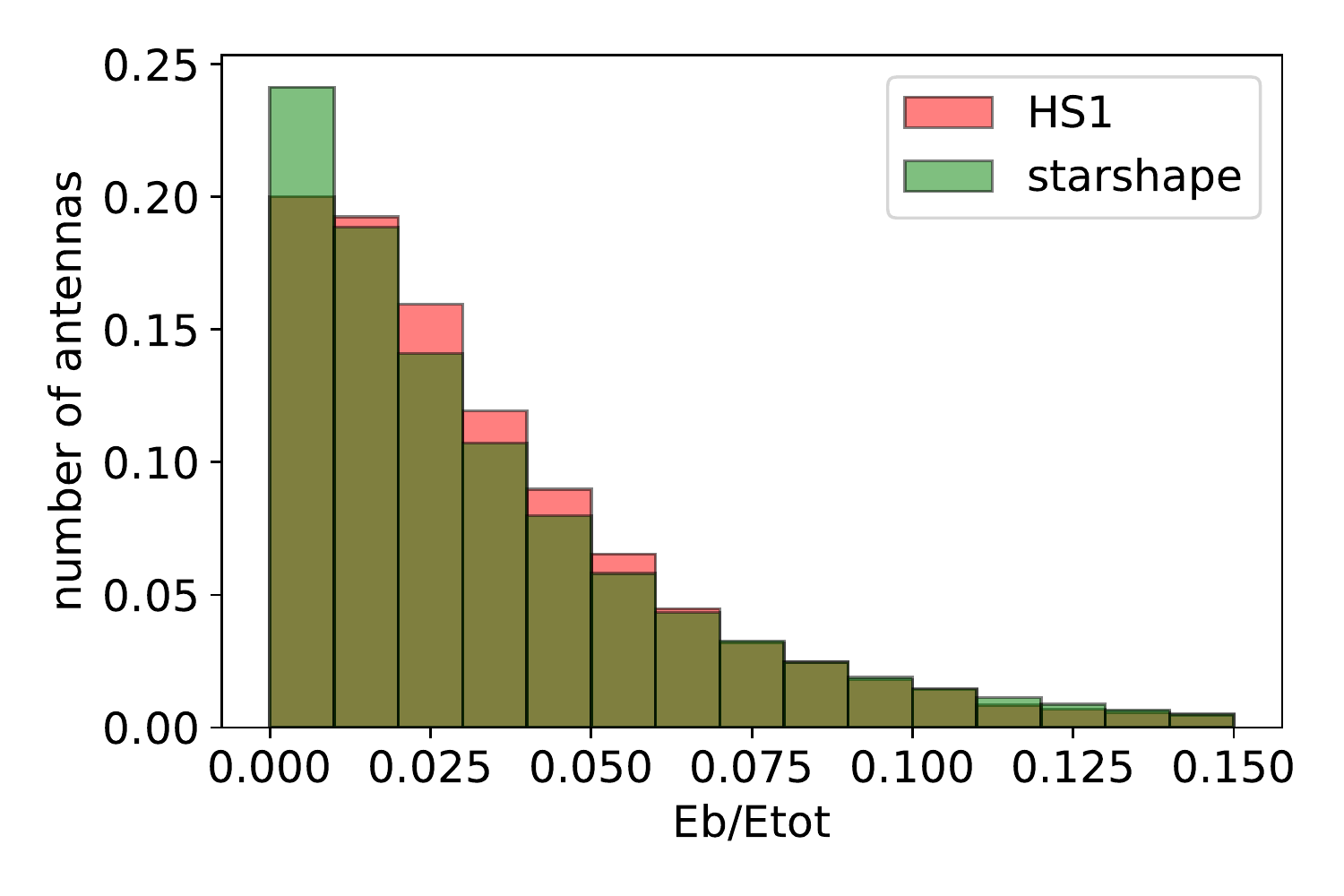}
\caption{ Normalised histogram of the $E_{b}/E_{\rm tot}$ ratio measured at each antenna for proton induced showers at the $3\sigma$ trigger threshold for either the HS1 either the star-shape layout.}\label{fig:Eb_ratio_HS1_vs_stshp}
\end{figure}

In Fig.~\ref{fig:Eb_ratio_HS1_vs_stshp}, we represent histograms of the $E_{b}/E_{\rm tot}$ ratio for proton induced showers at the $3\sigma$ trigger level comparing results from the star-shape and the HS1 layout. This reveals a particularly interesting feature: although the HS1 and the star-shape consist of very different layout, we retrieve a high similitude between their $E_{b}/E_{\rm tot}$ ratio distributions. Such similitude can be justified considering that due to the trigger condition on the amplitude, the antennas that trigger are likely to be the closest antennas to the Cerenkov cone independently of the layout configuration. This suggests that our criterion defined to perform autonomous radio detection seems to be almost independent of the  considered layout and could therefore be implemented for a wide variety of experiments dedicated to the radio detection of air-showers. 

\begin{figure*}[tb]
\includegraphics[width=0.50\linewidth]{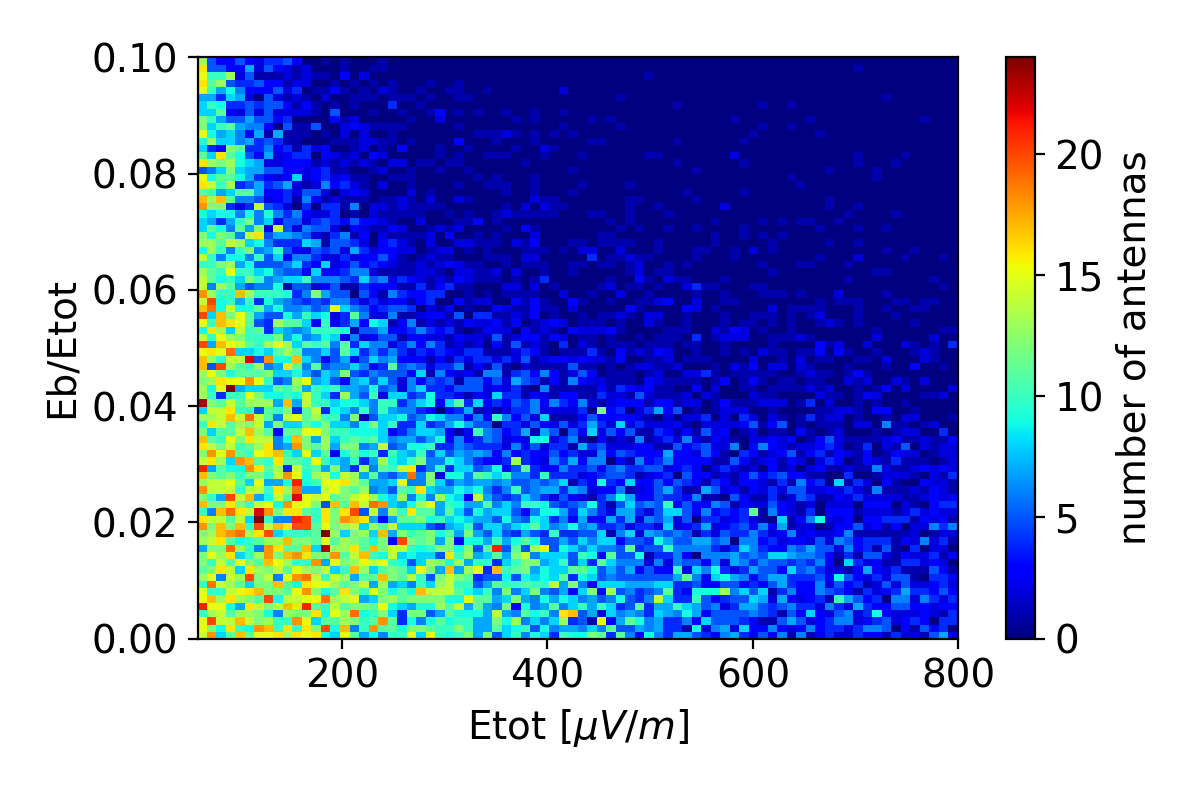}\hfill
\includegraphics[width=0.50\linewidth]{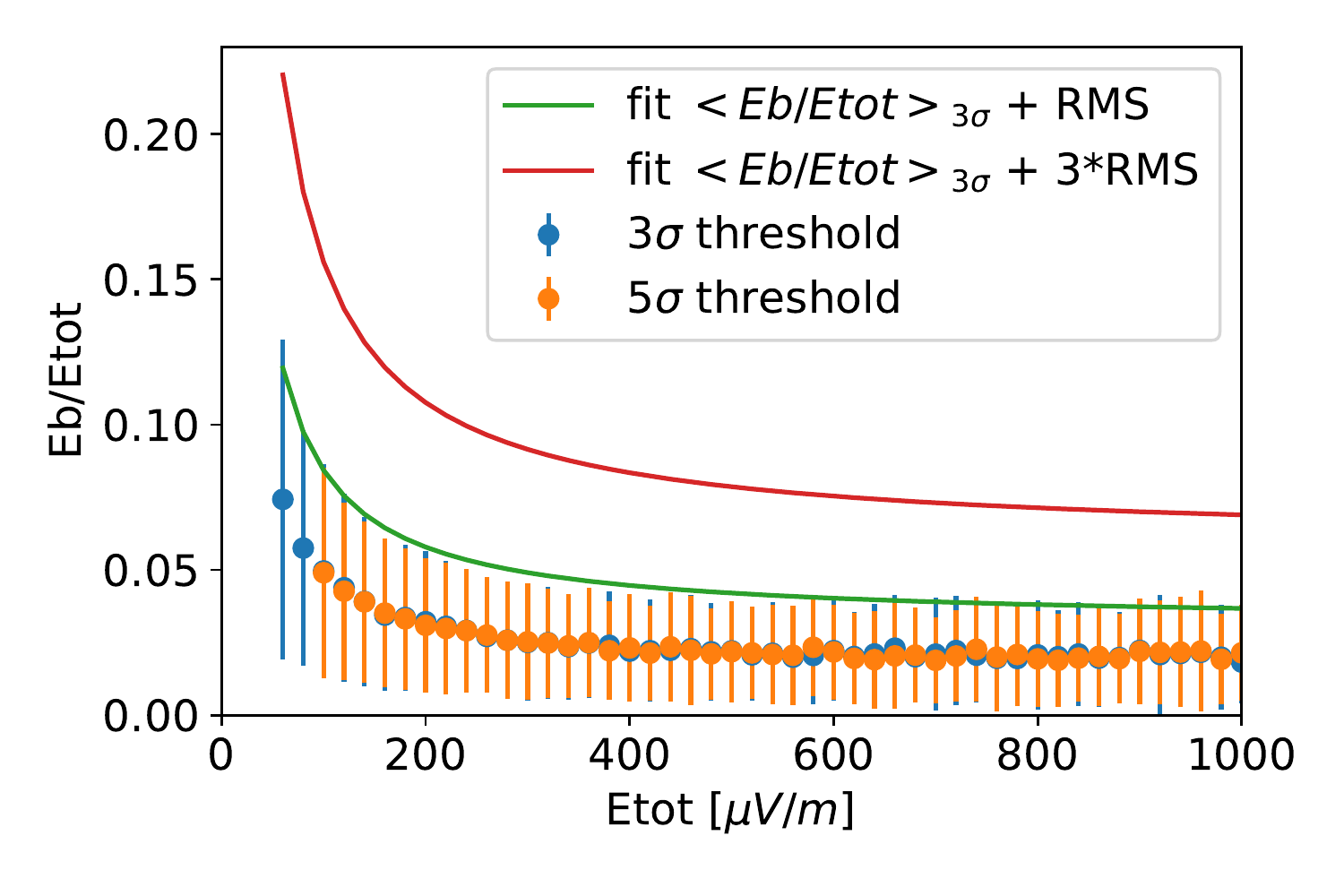}
\caption{$E_{b}/E_{\rm tot}$ ratio measured at each antenna for proton induced showers at the $3\sigma$ trigger threshold on the HS1 layout ({\it left}) and after averaging the ratio over different bins of the total amplitude for both $3\sigma$ and $5\sigma$ trigger thresholds ({\it right}).}\label{fig:Eb_vs_Etot_HS1}
\end{figure*}

Finally in Fig.~\ref{fig:Eb_vs_Etot_HS1} we represent the $E_{b}/E_{\rm tot}$ ratio as a function of the amplitude in a 2D plot at the $3\sigma$ trigger threshold (left panel) and after averaging the ratio over different bins of the total amplitude for both $3\sigma$ and $5\sigma$ trigger thresholds (right panel). Here again, we retrieve similar results with the ones of the star-shape layout, that is that the $E_{b}/E_{\rm tot}$ ratio decreases with increasing amplitude of the signal measured by the antennas, even though here our statistic is a bit reduced as seen on the left panel.  We can once again define a trigger condition $E_{b}/E_{\rm tot}< f(E_{\rm tot})$ where $f(E_{\rm tot})$ is obtained by fitting the points $\langle E_{b}/E_{\rm tot} \rangle + n\sigma_{E_{b}/E_{\rm tot}}$ with Eq.~\ref{eq:trigger_function}. For $n=1$, we find $a = 0.1775 \pm 0.0004$ and $b = 6 \pm 0.1$ with a $83\%$ trigger efficiency and for $n=3$  $a = 0.0330 \pm 0.0007$ and $b = 10.9 \pm 0.3$ with a $99\%$ trigger efficiency. This is in excellent agreement with the results of the star-shape simulation. 

\subsubsection{Polarized intensity}

\begin{figure}[tb]
\includegraphics[width=0.95\columnwidth]{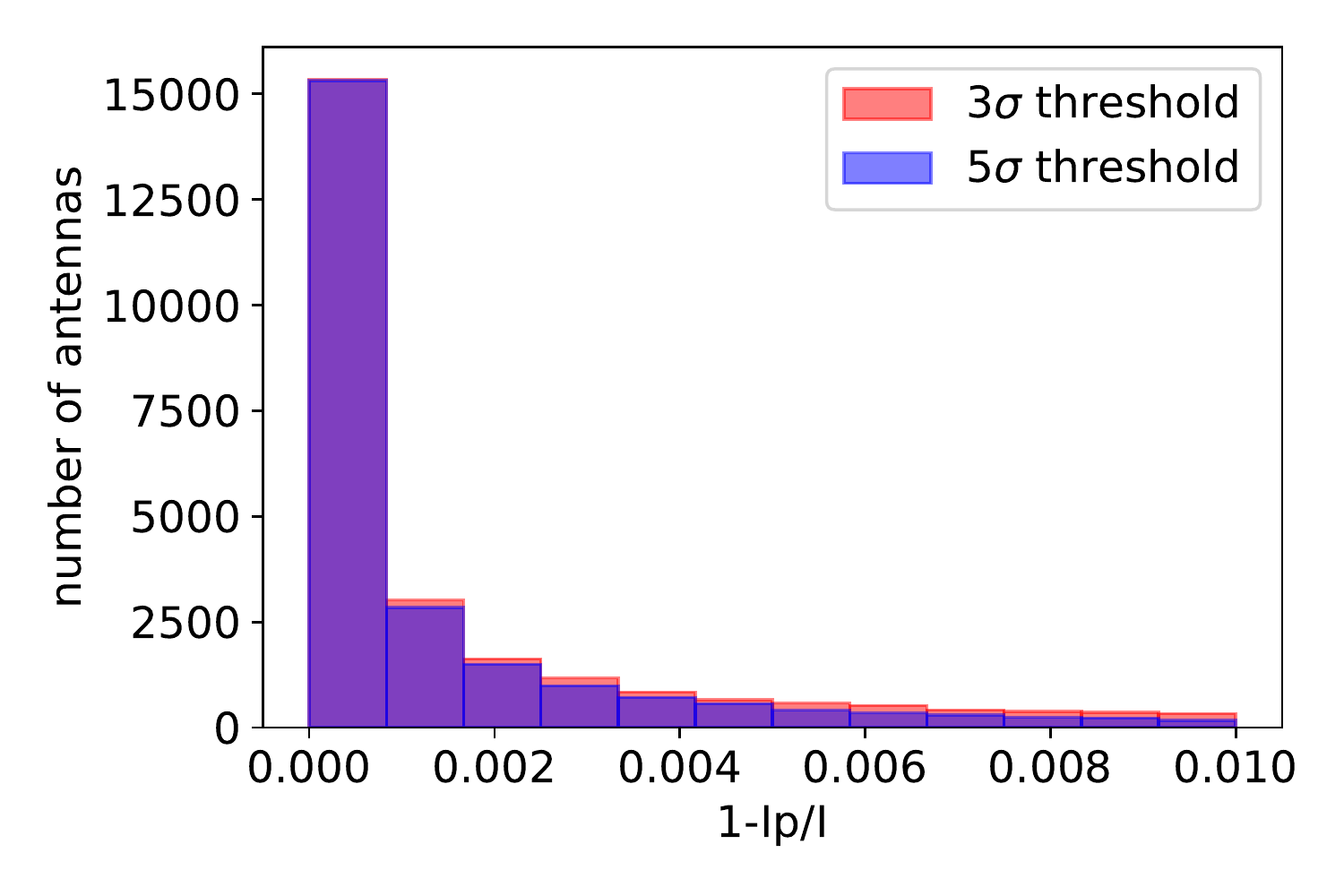}
\caption{ Histogram of $1 - I_{p}/I$ (where $I_{p}/I$ is the fraction of polarised intensity) measured at each antenna for proton induced showers on the HS1 layout at either a $3\sigma$ or  a $5\sigma$ trigger threshold.}\label{fig:Ip_ratio_HS1}
\end{figure}

In Fig.~\ref{fig:Ip_ratio_HS1}, we represent an histogram of $1-I_{p}/I$ computed at each antenna for simulations of proton induced showers on the HS1 layout for both trigger thresholds $3\sigma$ and $5\sigma$. It appears that the ratio is below 2\% for more than 90\% of the antennas at the $3\sigma$ trigger threshold and more than 98\% at the $5\sigma$ trigger threshold, which confirms that the signal from air-showers is highly polarized.\\

\section{Polarisation signatures of neutrino-induced air-showers}\label{section:neutrino}

Ultra-high energy neutrinos can also induce extensive air-showers. As these particles have a very low cross section, they need a dense environment to have a reasonable probability to interact. This can happen for example in the case of an up-going shower, where the neutrino interacts directly in the ground and decays into a tau particle. The tau particle can then emerge and decay in the atmosphere inducing an extensive air-shower. Antennas placed in a valley provide an alternative. A down-going tau neutrino can in this case interact within a mountain, produce a tau particle, which in turn escapes from the mountain and decays in the atmosphere, inducing an extensive air-shower as illustrated in Fig.~16 of~\cite{2020SCPMA..6319501A}.

Only Earth-skimming trajectories correspond to non-negligeable detection probabilities.  Neutrino showers are therefore expected to develop close to the ground, in denser environment than cosmic-ray showers. This signature should directly translate into a higher $a$-ratio. Indeed, we recall as can be seen from Eq.~\eqref{eq:scaling_ce_geo}, that a higher density should result in an increase of the charge excess (increase of $C_{x}$) and a decrease of the geomagnetic emission (decrease of $\langle v_{d} \rangle$).

 In this section, we study this trend for neutrino induced air-showers by comparing their $E_{b}/E_{\rm tot}$ ratios with the ones we obtained for cosmic-ray showers. For this purpose, we consider $1\,000$ simulations of neutrino showers, either down-going or up-going with zenith angles ranging from 85$\degree$ to  95$\degree$ ($\theta >90\degree$ for up-going showers) and various energies ranging from $0.1\,$EeV to $25\,$EeV. Simulation were performed on the HS1 layout as presented in Fig.~\ref{fig:HS1_layout} with different values for energy and direction of origin.

In Fig.~\ref{fig:Eb_ratio_neutrinos}, we represent the $E_{b}/E_{\rm tot}$ ratio for neutrino showers comparing a conservative and aggressive trigger threshold (left panel) and comparing neutrinos, proton and iron nuclei at the $5\sigma$ trigger threshold level (right panel). It appears clearly that the ratio is higher for neutrino than for cosmic-ray induced showers: the distribution peaks around 15\% for neutrino primaries and 2\% for cosmic-ray showers.  Still, we can also observe that for some antennas, even in the case of neutrino simulations, the measurement of the $E_{b}/E_{\rm tot}$ ratio can reach low values, close to 0. This is related to the radial signature of the charge excess, implying that its emission should drop to 0 for antennas close to the core position or for antennas along the { ${\bf v \times (v \times B)}$} baseline, as the charge excess and geomagnetic emission are both orthogonal to $B$ along this axis. This can also be seen from Eq.~\ref{eq:scaling_ce_geo}: for showers occurring in denser environments, the slope of the $a$-ratio will be steeper due to an enhancement of $C_{x}/ \langle v_{d} \rangle$, but the ratio still tends to 0 for $\omega = 0$.  

\begin{figure}[tb]
\includegraphics[width=0.95\columnwidth]{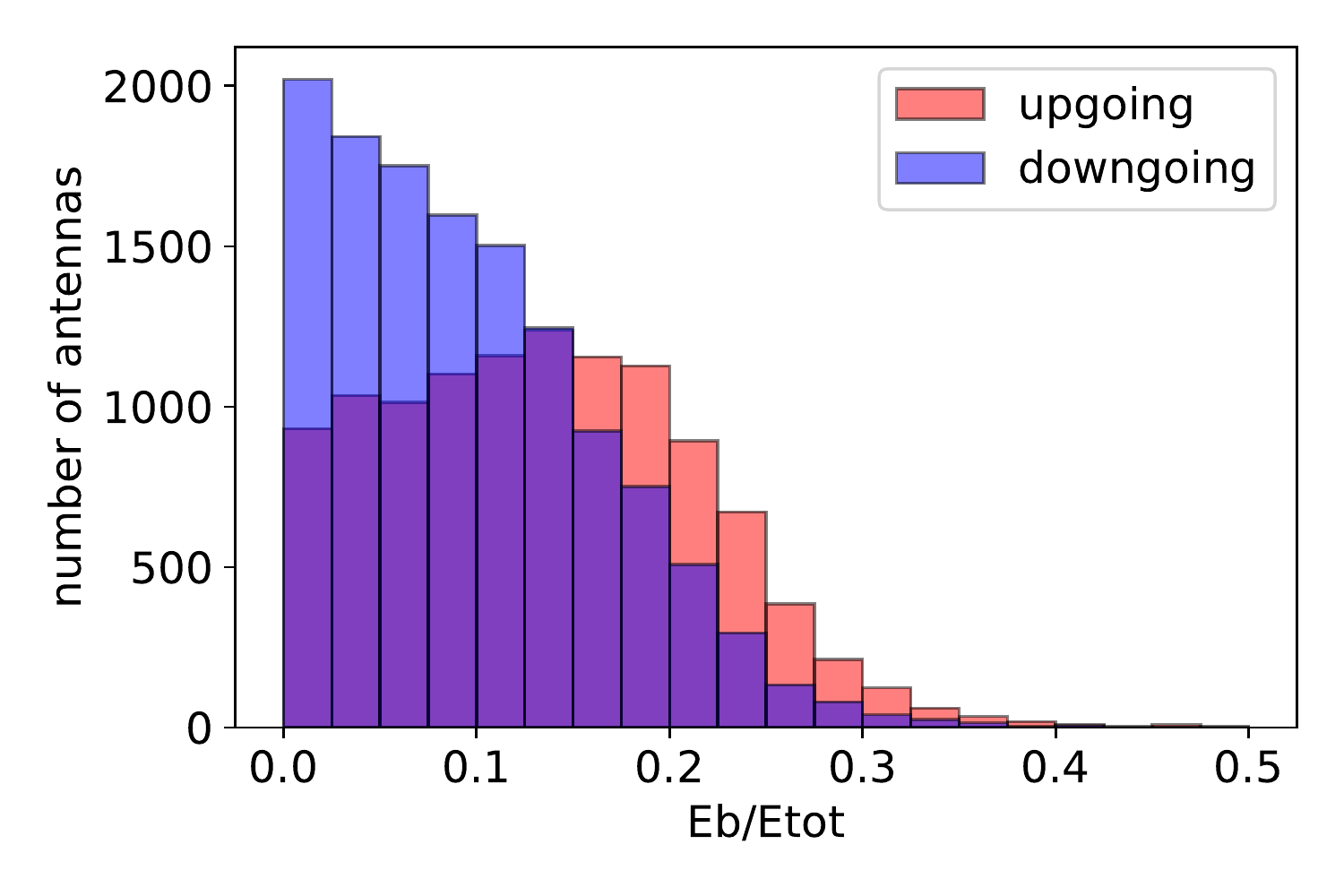}
\caption{ $E_{b}/E_{\rm tot}$ distribution comparing neutrino and cosmic-ray primaries on the Hotspot 1 layout at a $3\sigma$ trigger threshold.}\label{fig:neutrino_up_down}
\end{figure}

\begin{figure*}[tb]
\includegraphics[width=0.95\linewidth]{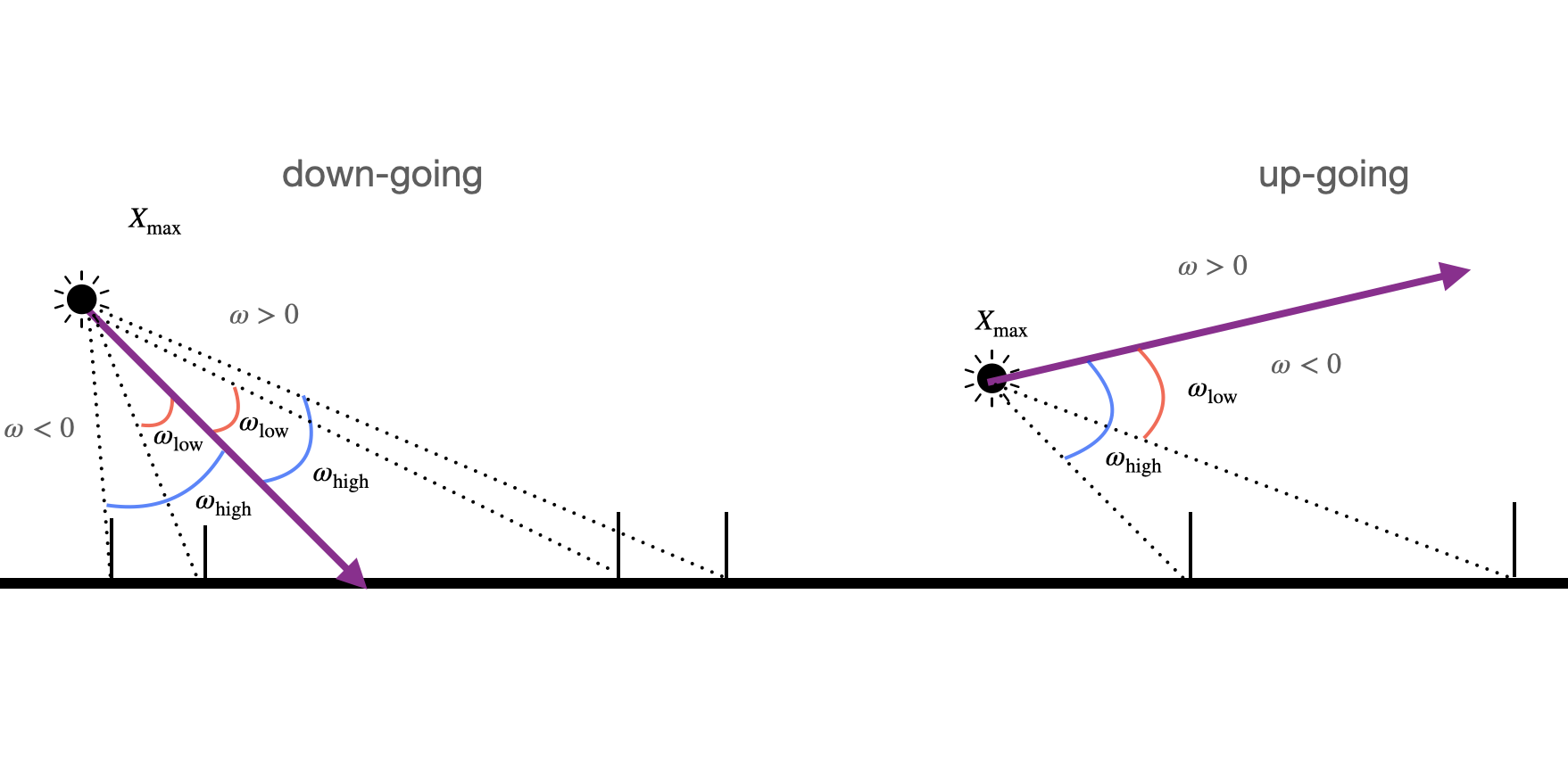}
\caption{Sketch of angular distance $\omega$ for down-going ({\it left}) and up-going ({\it right}) showers. Late
 antennas are more frequent for down-going than up-going showers, also the $X_{\rm max}$-antenna distance increases with the $\omega$ angle for late-antennas although it decreases with the $\omega$ angle for early-antennas.}\label{fig:sketch_updown}
\end{figure*}

\begin{figure*}[tb]
\includegraphics[width=0.49\linewidth]{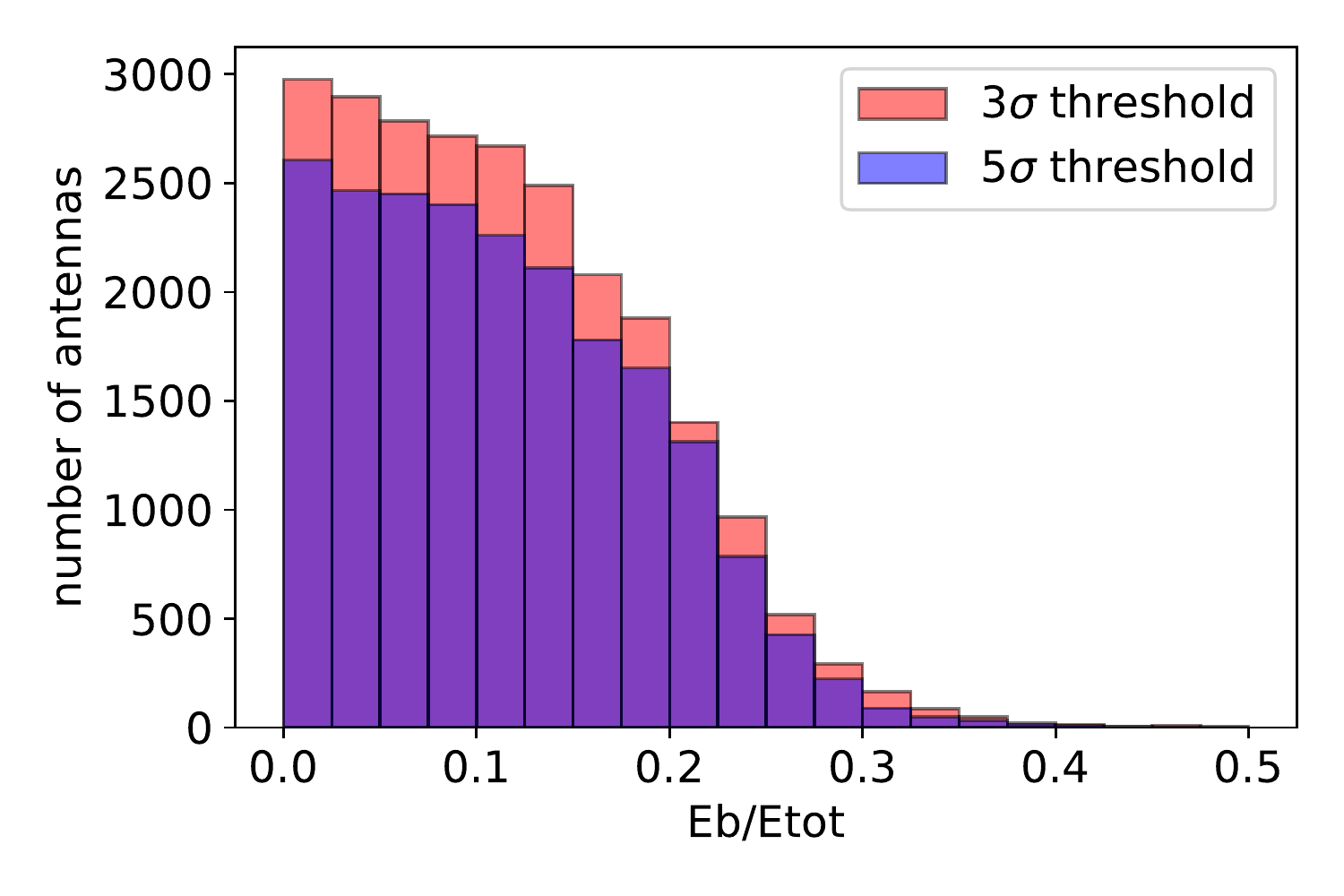}\hfill
\includegraphics[width=0.49\linewidth]{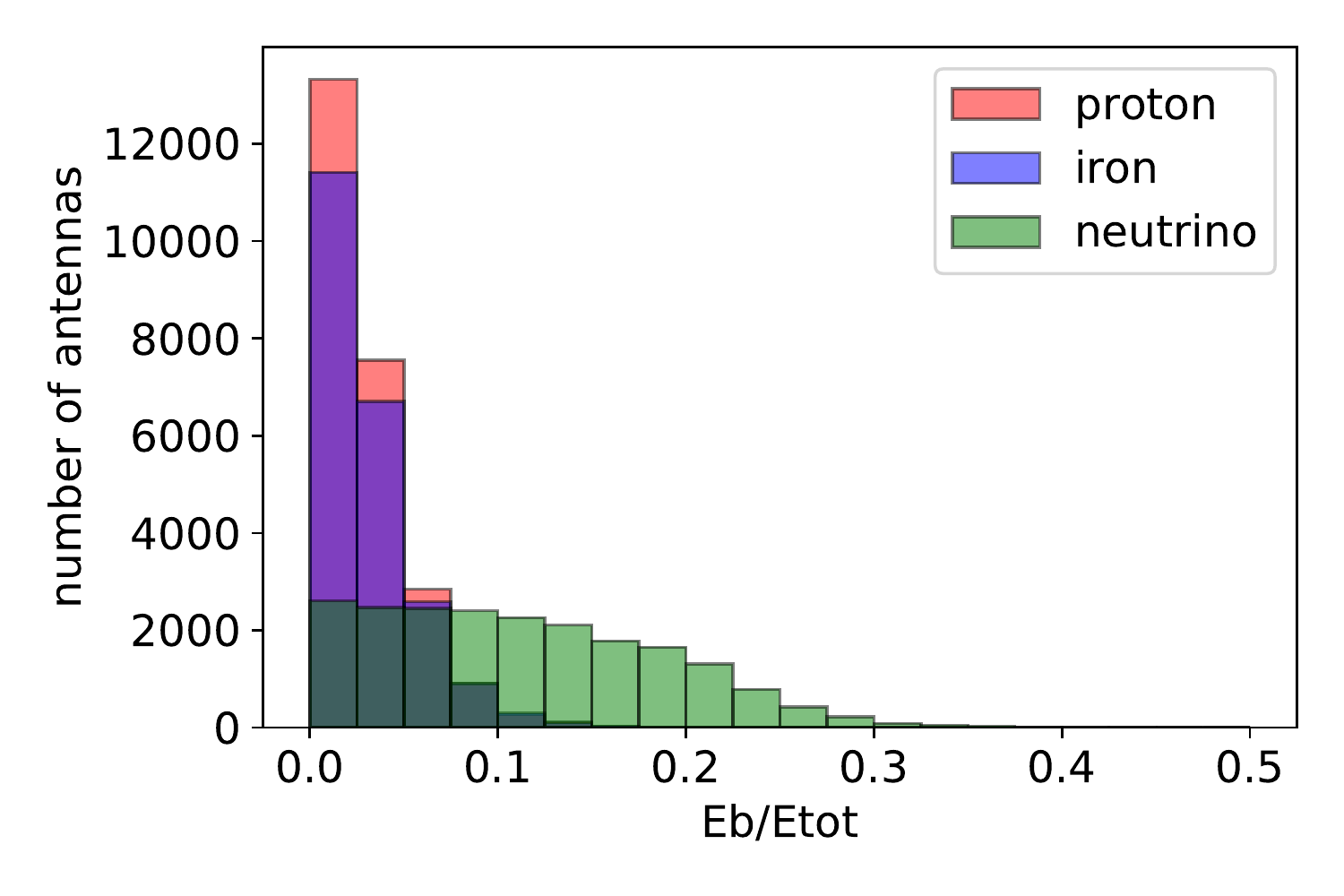}
\caption{Histograms of the $E_{b}/E_{\rm tot}$ ratio for neutrino simulations on the Hotspot 1 layout at different trigger thresholds ({\it left}) and comparing different primaries at a $5\sigma$ trigger threshold({\it right}).}\label{fig:Eb_ratio_neutrinos}
\end{figure*}

In Fig.~\ref{fig:neutrino_up_down}, we also represent the $E_{b}/E_{\rm tot}$ distribution for neutrino showers, but this time differentiating between up-going and down-going showers. It appears that the $E_{b}/E_{\rm tot}$ ratio is higher for up-going than down-going showers. This trend can be justified by geometrical considerations  as illustrated in Fig.~\ref{fig:sketch_updown}. The $\omega$ angle refers to the angular deviation between the shower axis and the direction that goes from $X_{\rm max}$ to a given antenna. The $\omega$ angle is counted positively or negatively following the trigonometric direction. On our sketch, it can be seen that for $\omega>0$, the $X_{\rm max}$-antenna distance increases with $|\omega|$ and that for $\omega<0$, the $X_{\rm max}$-antenna distance decreases as $|\omega|$ increases. Yet, for up-going showers, we mostly detect antennas with $\omega<0$, i.e., antennas that are closer to the emission point as $|\omega|$ increases. This bias towards antennas with $\omega<0$ for up-going showers implies that antennas with high $|\omega|$ values receive a less diluted signal and are more likely to trigger. This results in a higher $E_{b}/E_{\rm tot}$ ratio as was observed in Fig.~\ref{fig:ce_geo_w_starshape}.

\begin{figure}[tb]
\centering 
\includegraphics[width=0.95\columnwidth]{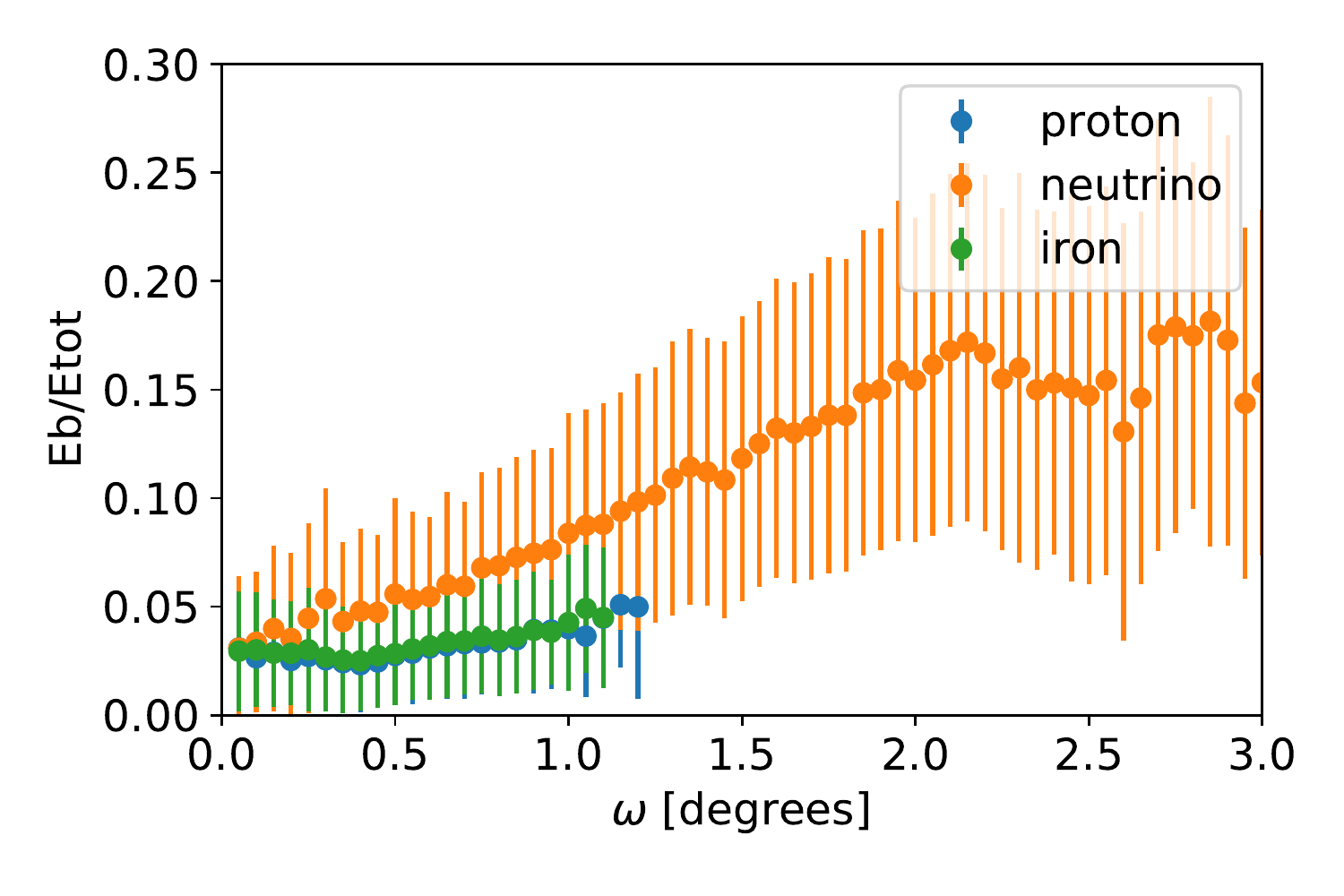}
\caption{$E_{b}/E_{\rm tot}$ ratio as a function of $\omega$ for proton, iron and neutrino induced showers at a $5\sigma$ trigger threshold, on the Hotspot 1 layout. }\label{fig:Eb_ratio_w_p_vs_iron_vs_nu}
\end{figure}

Finally in Fig.~\ref{fig:Eb_ratio_w_p_vs_iron_vs_nu}, we represent the comparison between the $E_{b}/E_{\rm tot}$ ratio as a function of $\omega$ for neutrino and cosmic-ray induced showers. It appears clearly that for a given $\omega$, values of the ratio will be higher for neutrinos than for cosmic-ray showers. As a consequence, we can infer that differences in the values of the ratio for neutrino and cosmic-ray showers as observed in Fig.~\ref{fig:Eb_ratio_neutrinos} are not (or at least not only) due to a different distribution of the antennas but density effects do play a role.

These results have several implications
\begin{enumerate}
    
\item Although the criteria defined in the previous section to perform autonomous radio detection are very efficient for cosmic-ray induced showers, it would have a much reduced efficiency for neutrino induced showers if applied as such. 
For experiments dedicated to the detection of neutrinos and cosmic-ray showers, the criteria defined in Section~\ref{section:trigger_thresholds} can be implemented only if a preliminary discrimination between neutrino and cosmic-ray showers can also be performed directly at the DAQ level.

\item It could be possible to discriminate between cosmic-ray and neutrinos induced showers using the $E_{b}/E_{\rm tot}$ ratio, as the ratio is expected to be distributed towards higher values for neutrinos induced showers. However, due to the radial signature of the charge excess, the $E_{b}/E_{\rm tot}$ ratio will drop to 0 for antennas close to the core or on the { ${\bf v \times (v \times B)}$} axis, implying that the discrimination between neutrino and cosmic-ray showers could not be performed at the antenna level for all the antennas but rather offline, averaging for example the ratio over the full array.

\item Finally, as the contribution of the charge excess emission is stronger for neutrino showers, it might be possible to use this charge excess signature to reconstruct air-shower parameters. For example, as illustrated in Fig.~\ref{fig:polarisation}, the charge excess emission points towards the air-shower core position and as shown in Fig.~\ref{fig:ce_geo_w_starshape}, the charge excess amplitude increases with the distance to the core. These are two strong signatures that could help us reconstruct the air-shower core position, allowing for angular reconstruction if the $X_{max}$ position can also be reconstructed, for example. This will be investigated in further dedicated studies.
\end{enumerate}

%%%%%%%%%%%%%%%%%%%%%%%%%%%%%%%%%%%%%%%%%%%
%%%%% SUMMARY
%%%%%%%%%%%%%%%%%%%%%%%%%%%%%%%%%%%%%%%%%%%
\section{Conclusion}

Based on numerical simulations of inclined cosmic-ray and neutrino air-showers, we have characterized polarisation signatures of the radio signal, that could serve to perform autonomous radio detection at the DAQ level. 

We focused on the two main contributions to the polarisation, namely the geomagnetic and the charge excess emissions. We have evidenced the clear predominance of the geomagnetic emission over the charge excess emission for inclined ($\theta >65\degree$) cosmic-ray air-showers. This implies that the total electric field from the radio emission of such air-showers should be in good approximation perpendicular to the direction of the local magnetic field. 

This specific feature can be exploited to reject noise and identify cosmic-ray triggered events. We have established trigger conditions directly at the DAQ level, based on the fraction of the total electric field projected along $\mathbf{B}$ ($E_{b}/E_{\rm tot}$). This quantity can be computed after reconstructing the electric field by deconvoluting the voltage from the antenna response, and is by construction sensitive to the geomagnetic emission polarisation. This of course requires first that a reconstruction of the shower direction is performed.

We tested our rejection method on both a star-shape and an experimental layout. For the experimental layout in particular, our results show that a uniform trigger condition on the $E_{b}/E_{\rm tot}$ ratio at 7\%
enables a rejection efficiency of $\approx 93\%$ for a favorable site (as defined in section \ref{section:ce_geo_dependencies}), and $\approx 72\%$ for a non-favorable one. The  trigger efficiency of this cut for cosmic-ray air-showers is $86\%$ (93\%) with a $3\sigma$ ($5\sigma$)-trigger threshold level on the total amplitude. Note that these values were obtained assuming an isotropic distribution in space and a gaussian distribution in amplitude of noise events and that the real noise environment of the target site should be used to get more accurate estimations. Moreover, we show that the accuracy of our trigger criteria for shower identification increases with the amplitude of the electric field. Hence implementing an amplitude-dependent trigger condition would strengthen our background rejection efficiency. As we expect the signal from air-showers to be highly polarised, the fraction on polarised intensity $I_{p}/I$ could also help perform shower identification. 

On the other hand, for neutrino-induced showers, we find that the higher contribution of the charge excess emission does not enable to perform shower identification using the same method as for cosmic-rays. However, the charge excess polarisation pattern should enable to perform an efficient reconstruction of the air-shower parameters as well as a discrimination between cosmic-ray and neutrino induced showers.

The method developed here is largely independent of the detector array configuration, which suggests that it could be applied to a large variety of experiments with high efficiency.

%%%%%%%%%%%%%%%%%%%%%%%%%%%%%%%%%%%%%%%%%%
%%%% ACKNOWLEGMENT
%%%%%%%%%%%%%%%%%%%%%%%%%%%%%%%%%%%%%%%%%%
\subsection*{Acknowledgments}

We thank Valentin Decoene, Eric Hivon, and Simon Prunet for very fruitful discussions. We are also grateful to Harm Schoorlemmer for his valuable input at the beginning of this project. This work is supported by the APACHE ANR grant (ANR-16-CE31- 0001) of the French Agence Nationale de la Recherche, the Programme National des Hautes Energies of CNRS/INSU with INP and IN2P3, co-funded by CEA and CNES, and the European Research Council under the European Unions Horizon 2020 research and innovation programme (grant agreement No. 805486).  Simulations were performed  using  the  computing  resources  at  the  CC-IN2P3 Computing Centre (Lyon/Villeurbanne – France), partnership between CNRS/IN2P3 and CEA/DSM/Irfu.   
%%%%%%%%%%%%%%%%%%%%%%%%%%%%%%%%%%%%%%%%%%%%%%%%%%%%%%
%%% BIBLIOGRAPHY
%%%%%%%%%%%%%%%%%%%%%%%%%%%%%%%%%%%%%%%%%%%%%%%%%%%%%%
%\bibliographystyle{JHEP}  % A&A bibliography style file (aa.bst)

\bibliographystyle{elsarticle-num} 
\bibliography{polarisation_air_shower}

\end{document}